\documentclass[manuscript]{aastex}

\usepackage{graphics}
\usepackage{amsmath}
\usepackage{epstopdf}
\usepackage{float}
\usepackage{color}
\catcode`\@=11

\newcommand{\gapprox}{\mathrel{\mathpalette\@versim>}}
\newcommand{\lapprox}{\mathrel{\mathpalette\@versim<}}
\newcommand{\propapprox}{\mathrel{\mathpalette\@versim\propto}}
\newcommand{\@versim}[2]
  {\lower3.1truept\vbox{\baselineskip0pt\lineskip0.5truept
\ialign{$\m@th#1\hfil##\hfil$\crcr#2\crcr\sim\crcr}}}
\catcode`\@=12

\begin{document}

\tolerance=5000             
\def\cl{\centerline}      
\parskip 8truept

\title{Magnetic-Field Amplification in the Thin X-ray Rims of SN1006
} 
\author{Sean M. Ressler$^{1}$, Satoru Katsuda$^{2}$, Stephen P. Reynolds$^{3}$, Knox S. Long$^{4}$, Robert Petre$^{5}$,
  Brian J. Williams$^{5}$, \& P. Frank Winkler$^{6}$} \affil{$^1$ Department of Physics, University of California Berkeley, Berkeley, CA 94720, USA  \\$^{2}$Institute of Space and Astronautical Science (ISAS), Japan Aerospace Exploration Agency (JAXA), 3-1-1 Yoshinodai, Chuo, Sagamihara, Kanagawa 252-5210, Japan \\ $^{3}$Physics
  Department, North Carolina State University, Raleigh, NC 27695, USA \\
  $^{4}$Space Telescope
  Science Institute, 3700 San Martin Drive, Baltimore, MD 21218, USA\\ 
$^{5}$NASA Goddard Space
  Flight Center, Greenbelt, MD 20771, USA\\ $^{6}$Department of Physics, Middlebury College, Middlebury, VT
  05753, USA}

\begin{abstract}

Several young supernova remnants (SNRs), including SN1006, emit
synchrotron X-rays in narrow filaments, hereafter thin rims, along
their periphery.  The widths of these rims imply 50 to $100 \mu$G fields
in the region immediately behind the shock, far larger than expected for
the interstellar medium compressed by unmodified shocks, assuming electron
radiative losses limit rim widths.  However, magnetic-field damping
could also produce thin rims.  Here we review the literature on rim
width calculations, summarizing the case for magnetic-field
amplification.  We extend these calculations to include an arbitrary
power-law dependence of the diffusion coefficient on energy, 
$D \propto E^{\mu}$. Loss-limited rim widths should shrink with
increasing photon energy, while magnetic-damping models
predict widths almost independent of photon energy.  We use these
results to analyze Chandra observations of SN 1006, in particular the
southwest limb.  We parameterize the full widths at half maximum (FWHM)
in terms of energy as FWHM $\propto E^{m_E}_{\gamma}$.  Filament widths
in SN1006 decrease with energy; $m_E \sim -0.3$ to $-0.8$, implying
magnetic field amplification by factors of 10 to 50, above the factor
of 4 expected in strong unmodified shocks.  For SN 1006, the rapid
shrinkage rules out magnetic damping models.  It also favors short
mean free paths (small diffusion coefficients) and strong
dependence of $D$ on energy ($\mu \ge 1$).

\end{abstract}


\section{Introduction}

Cosmic synchrotron sources, such as jets in active galactic nuclei,
radio halos and relics in clusters of galaxies, pulsar-wind nebulae,
and shell supernova remnants (SNRs), demonstrate the ubiquity of
power-law distributions of relativistic electrons.  Understanding the
origins of these fast particles is necessary to learn about these
objects' energy budgets and evolution.  The synchrotron flux density
emitted by a source depends roughly on the product of the energy
density of relativistic electrons $u_e$ and the magnetic field $u_B$,
but an independent determination of magnetic-field strengths in
synchrotron sources has proven elusive.  The minimum energy of a
synchrotron source occurs when the two energy densities are roughly
equal (``equipartition;'' actually, $u_e = (4/3)u_B$; e.g., Pacholczyk
1970).  However, it is not clear whether the unseen population of
relativistic protons should also be included, and if so, what the
proton-to-electron energy ratio should be.  Furthermore, there is no
obvious physical reason to expect equipartition.  The argument for
equipartition derives from attempts to explain extragalactic radio
sources in which the total energy budget is so large that it was of
interest to find a lower bound (Burbidge 1956).  However, many other
synchrotron sources, including SNRs, release a relatively small
fraction of their total energy content as synchrotron emission, so
could easily afford to be far from equipartition (in either
direction).

Although magnetic fields are not dynamically important in SNRs (e.g.,
Jun \& Jones 1999), their strength is critical in determining the
maximum energy to which particles can be accelerated.  For the
diffusive shock acceleration process (DSA; e.g., Blandford $\&$
Eichler 1987), the time $\tau(E)$ to accelerate particles to energy
$E$ depends on the diffusion coefficient $D$ and the shock velocity
$v_{\rm shock}$ by $\tau(E) \sim D/v_{\rm shock}^2$.  For relativistic
particles, $D = \lambda c/3$. Then in ``Bohm-like'' diffusion, where
the mean free path $\lambda$ is assumed proportional to the particle
gyroradius ($\lambda = \eta r_{\rm g} = \eta E/eB$), $\tau(E) \propto
1/B$, and higher magnetic fields result in more rapid acceleration and
higher maximum energies.  This is independent of which of several
competing mechanisms ultimately limits acceleration (finite time since
onset of acceleration, radiative losses, or escape). Note that for the
above description of the diffusion coefficient, taking $\eta = 1$ is
called Bohm diffusion or the ``Bohm limit.''  It corresponds to
$\lambda = r_g$, often assumed to be the shortest physically plausible
mean free path.  However, in a turbulent wave field, it is not clear
whether this is a true limit, or even what kind of average value for
the magnetic-field strength should be used to calculate $r_g$ (e.g.,
Reville \& Bell 2013).  

Largely by exclusion of competing hypotheses, Galactic cosmic ray
acceleration is now widely attributed to SNR shocks.  The consensus is
that SNRs can accelerate particles up to the ``knee,'' the slight
inflection and steepening around 3 PeV ($3 \times 10^{15}$ eV).  (No
plausible version of SNR-based DSA produces the maximum energies
observed in cosmic rays of above $10^{19}$ eV [e.g., Abraham et
  al.~2008], for which an extragalactic origin is presumed.)  However,
since the work of Lagage \& Cesarsky (1983), it has been clear that
typical estimates of magnetic-field strengths behind SNR shocks of a
few $\mu$Gauss (the mean interstellar magnetic field multiplied by the
shock compression ratio, $r$, taken to be 4 for strong nonrelativistic
shocks), result in maximum energies that fall short of the ``knee'' by
an order of magnitude or more.  These estimates of $B$ are based on
measurements of the interstellar magnetic field strength within a few
kpc of the Sun, which is about 2 -- 3 $\mu$Gauss (Lyne \& Smith 1989).
Simple compression in a strong shock with adiabatic index $\gamma =
5/3$ (i.e., unmodified by cosmic rays) would produce downstream values
larger by a factor of up to 4 (no amplification if the shock velocity
is parallel to the field, a factor of 4 increase if perpendicular).
Thus magnetic fields larger than about 12 $\mu$Gauss require an
additional process of amplification.  Independent methods of
estimating interstellar magnetic-field strengths, such as Zeeman
splitting in molecular lines, are not relevant to SNR environments.

Thus the plausibility of models in which SNR shocks produce Galactic
cosmic rays up to the ``knee'' may depend on observational
determinations of the post-shock magnetic field strength.
Fortunately, the realization that young SNRs can produce synchrotron
emission into the X-ray band has made available a new, potentially
powerful method for such determinations.  Several young SNRs show thin
(a very small fraction of their diameter), synchrotron-emitting
filaments along their edges. The widths of these rims in the radial
direction have been used to infer estimates for the post-shock
magnetic field strength and presented as evidence for significant
magnetic field amplification by strong shocks, as we shall review in
detail below.  The complexity of the calculations used to infer the
magnetic field magnitude varies significantly from simple analytic
approximations to detailed numerical calculations, but the consensus
from these studies is that field amplification well beyond a factor of
four is required to explain the X-ray observations.

Thin synchrotron rims present a well-defined problem.  While
synchrotron emissivity may suddenly turn on at the shock front due to
particle acceleration and magnetic-field compression (or
amplification) there, turning it off again only about a tenth of a
parsec downstream is not so simple: one must either eliminate the
radiating particles or the magnetic field.  Both possibilities have
been suggested. Radiative energy losses as particles advect and
diffuse downstream will eventually lower electron energies below the
level at which synchrotron X-rays can be produced.  More rapid
synchrotron losses for higher electron energies then predict that rims
will become thinner at higher photon energies.  However, it is also
possible that magnetic fields somehow decay behind the shock, with a
length scale (almost) independent of electron energy, predicting rims
whose thickness is relatively constant with photon energy.

In the energy-loss scenario, the rate at which rims become thinner as
observing energy rises depends on electron transport.  If electrons
are simply advected downstream, rim widths $l$ depend only on the
magnetic-field strength, and drop quite rapidly with increasing photon
energy, $l \propto (h \nu)^{-1/2}$ in the simplest approximation, as
we describe below.  However, diffusion allows electrons of the same
energy to spread out spatially, diluting this effect somewhat and
predicting slower drops in rim widths with energy.  Measuring the rim
widths at several different photon energies is thus key to
discriminating among models.

The remnant of the Type Ia Supernova of AD 1006 is well suited for
this analysis, as its large angular size coupled with \emph{Chandra}'s
high spatial resolution allows accurate measurements of radial
profiles of the filaments: the remnant radius of about $15'$
corresponds to over 1800 {\sl Chandra} ACIS pixels. Furthermore, the
shock speeds in the synchrotron-dominated northeast (NE) and southwest
(SW) edges are about 5000 km s$^{-1}$, as measured from their proper
motion (Katsuda et al.~2009, Winkler et al.~2014), and the SNR has
been detected as a TeV source (Acero et al.~2010), suggesting that
SN1006 produces very high-energy cosmic rays.

Our purpose here is to consider theories of particle diffusion and
magnetic-field amplification in the light of new deep observations of
SN1006 made with \emph{Chandra}.  In Section 2 (summarized in
Table~\ref{tab:rev}), we review earlier work on filament calculations
and the evidence for field amplification to establish a firm
background for the new work presented here.  In Section 3, we
generalize previous work by allowing different energy dependence of
the diffusion coefficient from Bohm-like, including Kolmogorov and
Kraichnan-type diffusion, among others.  We first neglect any cut-off
in the electron spectrum and calculate model profiles and their energy
dependence for the loss-limited (Section 3.1) and magnetically damped
(Section 3.2) scenarios. We then add the effects of an electron
cut-off energy (Section 3.3), and examine the effects of making the
$\delta$-function approximation for the emissivity (Section 3.4).
In Section 4, we describe measurements of the widths of the nonthermal
filaments in SN1006, including for the first time the SW region,
making use of new high-resolution \emph{Chandra} measurements.  We
find that rim widths decrease with increasing photon energy, quite
rapidly in some cases.  We use these measurements to
constrain both the post-shock magnetic field and diffusion
coefficient.  The general narrowing of rims eliminates the
magnetic-damping model for rim widths; quantitatively, we find values
of post-shock magnetic field of 70 -- 200 $\mu$G, comparable to those
obtained in earlier work.  However, the rapidity of rim shrinkage
suggests that diffusion mean free paths in some areas are quite small,
perhaps less than the gyroradius.  We discuss these results in Section
5, including reviewing theoretical and observational work on sub-Bohm
diffusion and implications for particle acceleration to high energies.
We summarize our conclusions in Section 6.  Finally, in the appendix,
we offer a list of the results of our experience in applying the
various rim models to observations, as a guide for potential future
investigations.


\section{Previous Work}

Prior work bifurcates into two eras, an earlier one in which it was
assumed that the only influence on filament shapes was synchrotron
losses, followed by one beginning in 2005 when additional effects such
as magnetic field damping began to be introduced.
We consider first the former case; decay of magnetic field is
considered in Section 2.2.

\subsection{Loss-limited models}

Before 2005, it was universally assumed that the shapes of nonthermal
X-ray filaments observed in SNRs were due to synchrotron losses by
high energy electrons (see the many references described below and in
Table~\ref{tab:rev}). The idea was that an electron could only travel
a certain distance before losing enough energy that its radiation
dropped below the X-ray band. This distance is determined by two
competing transport mechanisms: advection (bulk motion of plasma) and diffusion (random motion of electrons on the scale of gyroradii). Considered
separately, one can obtain simple expressions for the appropriate
length scale for each in terms of the diffusion coefficient, $D$, the
downstream plasma speed in the shock frame, $v_{\rm d}$, and the
synchrotron cooling time, $\tau_{\rm synch}$. We can estimate
$\tau_{\rm synch} = 1/(bB^{2}E)$, where $b = 1.57\times 10^{-3}$ in
cgs, from the relation $\dot{E}\propto E^{2}$.  Thus, for $v_{\rm
  d}=v_{\rm shock}/4$, given by the Rankine-Hugoniot conditions for a
strong adiabatic shock, we have an advective length of $l_{\rm ad}
\approx v_{\rm d}\tau_{\rm synch}=(v_{\rm d})/(bB^{2}E)$. For a
diffusion coefficient that is taken as a constant multiple, $\eta$, of
the Bohm value $D = \eta C_{\rm d}E/B$, where $C_{\rm d} \equiv
c/(3e)$, we arrive at a diffusive length of $l_{\rm diff} \approx
\sqrt{D\tau_{\rm synch}}=\sqrt{D/(bB^{2}E)}$. For the values of the
constants $b$, $C_{\rm d}$, $c_{\rm m}$, and $c_1$ used here and
throughout, see Table~\ref{tab:const}. Now, an electron of energy $E$
in a magnetic field radiates primarily at the frequency $\nu_m =
c_{\rm m}E^{2}B$ so that the advection and diffusion lengths as a
function of frequency are
\begin{equation}
l_{\rm ad} = \frac{v_d\sqrt{c_{m}}}{b}B^{-3/2}\nu_m^{-1/2}
\end{equation}
and
\begin{equation}
l_{\rm diff} = \sqrt{\frac{\eta C_{\rm d}}{b}}B^{-3/2}.
\end{equation}
The approximation that the electron radiates all its energy at
$\nu_m$ is called the delta-function approximation.
\textbf{The important result here is that $l_{\rm ad}$ varies as
$\nu^{-1/2}$ while $l_{\rm diff}$ is independent of frequency.} Thus
above some critical energy, $E_{c}$, and an associated photon
frequency, $\nu_{c}$, electrons will be able to diffuse further in a
loss time than they could advect, and electron diffusion will become
the dominant method of transport.  $E_{\rm c}$ is found simply by equating
the expressions for $l_{\rm ad}$ and $l_{\rm diff}$:
\begin{eqnarray}
E_{\rm c} = \frac{v_{\rm d}}{\sqrt{\eta C_{\rm d}bB}} \approx 69.12 \ {\rm ergs} 
  \left(\frac{v_{\rm d}}{1250 \ {\rm km\ s}^{-1}}\right)
  \left(\frac{B}{100\ \mu{\rm G}}\right)^{-1/2} \eta^{-1/2}\\
h\nu_{\rm c} = \frac{c_{m}v_{\rm d}^{2}}{\eta C_{\rm d}b} \approx 3.61\ {\rm keV} 
  \left(\frac{v_{\rm d}}{1250\ {\rm km\ s}^{-1}}\right)^{2}\frac{1}{\eta}.
\end{eqnarray} 
(We have taken $v_d = v_{\rm shock}/4 = 1250$ km s$^{-1}$, assuming no
shock modification by cosmic rays.)  Near this photon energy, both
advection and diffusion are important.  This simple approach was taken
by Ballet (2006), Bamba et al.~(2003), and Yamazaki et al.~(2004) to
infer magnetic field strengths of 14--87 $\mu$G in SN 1006. Vink $\&$
Laming (2003) used a similar technique for Cas A, and estimated $B$ to
be $\sim$ 100 $\mu$G.

Parizot et al. (2006) adopted a somewhat more sophisticated approach,
combining both processes in the steady state form of the
one-dimensional transport equation to solve for the post-shock
electron distribution $f(p,x)$ (V\"olk et al.  1981):
\begin{equation}
v\frac{\partial f}{\partial x} - D\frac{\partial^{2} f}{\partial
  x^{2}} + \frac{f}{\tau_{\rm synch}}=0,
\end{equation} 
where the loss term $f/(\tau_{\rm synch})$ assumes that an electron
maintains constant energy as it travels away from the injection site
until a catastrophic dump at time $\tau_{\rm synch}$.  Here the shock
is at $x = 0$ and $x > 0$ is the distance downstream.
The solution to this equation
is $f(p,x) \propto e^{-|x|/a}$, with the scale length $a$ given by:

\begin{equation}
a =\frac{2D/v_{\rm d}}{\sqrt{1+\frac{4D}{v_{\rm d}^{2}\tau_{\rm synch}}}-1}
\end{equation}
(Berezkho \& V\"olk 2004, Parizot et al. 2006).
More explicitly, in terms of electron energy $E$ the scale length is
\begin{equation}
a =\frac{2\eta C_{\rm d}E/Bv_{\rm d}} {\sqrt{1+\frac{4b \eta C_{\rm
        d}E^{2}B}{v_{\rm d}^{2}}}-1}.
\end{equation}
At a given observation frequency $\nu$, $E$ will depend on the magnetic
field $B$, by $E = \sqrt{\nu/(c_{m}B})$, so
\begin{equation}
a=\frac{2 \eta C_{\rm d}c_{m}^{-1/2}\nu^{1/2}B^{-3/2}/v_{\rm d}}
  {\sqrt{1+\frac{4b\eta C_{\rm d}\nu}{c_{m}v_{\rm d}^{2}}}-1}.
\end{equation}  
In order to estimate the strength of the  magnetic field, we  invert
this expression so that it is a function of observables:
\begin{equation}
B = \left(\frac{ac_{m}^{1/2}v_{\rm d}\left(\sqrt{1+\frac{4b\eta C_{\rm d}\nu}
  {c_{m}v_{\rm d}^{2}}}-1\right)}{2\eta C_{\rm d}\nu^{1/2}}\right)^{-2/3}.
\end{equation}  

In this equation $a$ is the scale length of the electron
distribution, whereas we observe the scale length of the emitted
synchrotron intensity, including line-of-sight projection effects.  In
the $\delta$-function approximation of the synchrotron emissivity,
$j_{\nu} \propto \sqrt{\nu B}f(E,x)$ ($E = pc$ for relativistic
electrons), and the radial intensity for a spherical shock is
\begin{equation}
I_{\nu}(r) =
2\int\limits^{\sqrt{r_{s}^{2}-r^{2}}}_{0}j_{\nu}\left(r_{s}-\sqrt{s^{2}+r^{2}}\right)ds.
\end{equation}
Here $r$ is the sky-plane radius ($r_s$ the shock radius), and $s$ is
the line-of-sight coordinate.  The resulting profile will have a FWHM
$= \beta a$, $\beta$ a projection factor.  
Ballet (2006) showed that
in the case of a purely exponential (in space) electron distribution
and for a spherical source, the result of this integral will give
$\beta = 4.6$, that is, a filament with a Full Width at Half Maximum
(FWHM) of $4.6a$. Thus in terms of the observed filament width,
$w_{\rm obs}$, we have
\begin{equation}
B = \left(\frac{w_{obs}c_{m}^{1/2}v_{\rm d}\left(\sqrt{1+\frac{4b\eta C_{\rm d}\nu}{c_{m}v_{\rm d}^{2}}}-1\right)}{2 \beta \eta C_{\rm d}\nu^{1/2}}\right)^{-2/3}.
\end{equation}
Using this result, the post-shock magnetic field strength in the NE
rim of SN1006 was estimated to be around 91-110 $\mu$G for $w_{obs} =
20''$, an amplification of roughly 30-37 for an ambient 3 $\mu$G field
(Parizot et al. 2006).  While those authors did not make use of
the fact, we note that the inferred value of $B$ depends on 
observing frequency.  

It should be noted that the projection factor $\beta = 4.6$ is
entirely dependent on the exponential form of the synchrotron
emissivity given from the solution of (5), which may not be valid, as
well as on the assumption of exact sphericity. This is an important caveat, as the width of the rims
scales as $B^{-3/2}$ (from equation (8)), so, since the width is inversely proportional to the projection factor, $\beta$, the above estimates for the post-shock magnetic field strength are proportional to $\beta^{2/3}$.    
In our later calculations, however, we do not assume a
simple constant projection factor and perform the full numerical
line-of-sight integration.

Finally, the most sophisticated synchrotron-loss based models of
Berezhko et al.~(2003), Berezhko $\&$ V\"olk (2004),
Cassam-Chena\"\i\ et al.~(2007), Morlino et al.~(2010) and Rettig $\&$
Pohl (2012) use an electron distribution obtained by solving the
continuous energy loss convection-diffusion equation (properly, the
advection-diffusion equation):
\begin{equation}
v\frac{\partial f}{\partial x}-\frac{\partial}{\partial x}\left(D
\frac{\partial f}{\partial x}\right)- \frac{\partial}{\partial
  E}\left(bB^{2}E^{2}f\right) = K_{0}E^{-s}e^{-E/E_{\rm cut}}\delta(x),
\end{equation}
where it is assumed that electrons are injected at the shock and
follow a power-law energy distribution with an exponential cut-off:
($N(E) \propto E^{-s} e^{E/E_{\rm cut}}$, where $s = 2.2$ -- the value
appropriate for SN 1006).  The electron distribution obtained from
solving this equation is convolved with the single-particle emissivity
and then integrated along lines of sight (see equation 10) to 
compute radial intensity profiles.  The magnetic-field strength for SN
1006 predicted using this method is in the range of 90-130 $\mu$G, an
amplification of roughly 30--43 for an ambient field of 3 $\mu$G
(Berezhko et al.~2003, Morlino et al.~2010, Rettig $\&$ Pohl 2012).

\begin{table}[h]
\caption{Previous Magnetic Field Strength Estimates for SN1006}
Note: $B_{0} \equiv B$ directly behind the shock
\centering
\scalebox{.65}{
\begin{tabular}{ c c c c}
\hline \hline
Paper & Technique &  SN1006 $B_{0}$ estimate & Amplification Factor\\  & & & (For ISM Field of 3 $\mu$G) \\ [.5ex]
Araya et al.~(2010) & Catastrophic Dump C-D equation & - & - \\
Ballet (2006) & Equated $l_{\rm diff}$ to rim sizes& 87 $\mu$ G & 29 \\
Bamba et al.~(2003) & Equated max$(l_{\rm diff}, l_{\rm ad})$ to rim sizes & - & -\\
Berezhko et al.~(2003) & Time dependent continuous loss C-D equation &
$\sim$ 100$\mu$G & 33\\
Berezhko $\&$ V\"olk (2004)&  Time dependent continuous loss C-D equation & - & -\\
Cassam Chena\"\i\ et al.~(2007) & CR modified numerical solution to C-D
equation & -  & -\\
Morlino et al.~(2010) & Nonlinear DSA Model Fit & 90 $\mu$G & 30\\
Parizot et al.~(2006)&Catastrophic Dump C-D equation +
$\delta$-function & 91-110 $\mu$G & 30-37\\
Rettig $\&$ Pohl (2012) & Continuous loss C-D equation & 130 $\mu$G
(loss limited) & 43 \\
  & & $\sim$65 $\mu$G (B-limited) & 22\\
Vink $\&$ Laming (2003) & Equated max$(l_{\rm diff}, l_{\rm ad})$ to rim sizes & - & - \\
Yamazaki et al.~(2004) & Equated max$(l_{\rm diff}, l_{\rm ad})$ to rim sizes
 & 14-85 $\mu$G & 5-28\\

\hline

\hline
\end{tabular}}
\label{tab:rev}
\end{table}

\subsection{Energy Dependence of the Filament Widths}

In almost all previous calculations, the energy dependence of the
filament width was ignored and profiles were fit at a single photon
energy. One exception to this is the work of Araya et al.~(2010) in
their analysis of the shapes of the rims of Cas A.  Interestingly,
while Araya et al.  found no significant energy dependence in the rim
widths between the energy ranges 3--6 keV and 6--10 keV, they did
report a small but non-negligible difference between widths at 0.3--2
and 3--6 keV.  However, they did not use this result as a parameter
constraint. Here we show that this dependence has important
physical consequences for parameter estimation.

In all the models we shall consider, the diffusion coefficient rises
with energy.  This means that electrons with lower energies will stay
closer to their original fluid element while those with higher
energies move about more freely. Specifically, as can be seen from
Equation 6, for $\nu \gg \nu_{\rm c}$, $a \approx l_{\rm diff}$, while
for $\nu \ll \nu_{\rm c}$, $a \approx l_{\rm ad}$.  This behavior
shows up in how the width, $a$, varies with energy, which we can parameterize as $a \propto E_{\gamma}^{m_{\rm E}}$ for a photon energy of $E_\gamma \equiv hv$.  Written in this way, we have
\begin{equation}
m_{E} = -\frac{1}{2}
  \left(1-\frac{4D/(v_{\rm d}^{2}\tau)}
     {1+4D/(v_{\rm d}^{2}\tau)-\sqrt{1+4D/(v_{\rm d}^{2}\tau)}}\right),
\end{equation}
where $4D/(v_{\rm d}^{2}\tau) \propto E^{2}B \propto \nu$, with the
last proportion coming from the $\delta$-function approximation,
meaning that $m_{E}$ is independent of magnetic field strength. It is
also clear from equation (13) that $m_{\rm E}$ will go from $-1/2
\rightarrow 0$, or in other words that the scale length, $a$, will go
from $a\propto E_{\gamma}^{-1/2} \rightarrow a \propto
E_{\gamma}^{0}$, as $\nu$ goes from $0 \rightarrow \infty$ (for $D
\propto E$, or in our later notation, $\mu = 1$).

\subsection{Magnetic-Field Damping}

In 2005 Pohl, Yan, \& Lazarian introduced a more
sophisticated approach, which suggested that claims of strong field
amplification might be premature.  They proposed several processes
that could lead to an exponentially decaying magnetic field, as well
as account for the narrow filamentation. In this case, rim profiles
would reflect the spatial distribution of the magnetic field.
Unfortunately, there are no simple predictions for the magnetic-field
damping length (or detailed spatial dependence).  There are a variety
of physically possible damping mechanisms, so the damping length is a
free parameter in models of this type (although its dependence on the
immediate post-shock value of $B$ can be preserved).

Cassam-Chena\"\i\ et al.(2007) used this idea to fit intensity
profiles of the filaments in Tycho's SNR, employing a hydrodynamics
code that included cosmic-ray shock modification (increased
compression ratios due to energetically important particles becoming
relativistic and/or escaping).  They generated model profiles assuming
a magnetic-field profile with exponential damping, similar to our
expression in Section 3.4 below.  By incorporating radio observations,
they concluded that the synchrotron loss-limited model provides a
slightly better fit than the magnetically damped model, though neither
completely reproduces the radio profiles. More recently, Rettig and
Pohl (2012) followed up by probing the observational consequences of
both a magnetically damped model and a constant field model by using
differences in the spectral index between the emission at the rim peak
(i.e. the emission from the shock front to a FWHM distance away)and in
the ``plateau'' (i.e. the emission from regions beyond the
FWHM). Their magnetic-field estimates for both models still favor
$\gtrsim$ 60 $\mu$G for SN1006.

Marcowith \& Casse (2010) performed detailed calculations to
investigate the magnetic-damping model, studying the amplification
process due to linear and nonlinear cosmic-ray streaming
instabilities, and identifying processes to damp the turbulent
magnetic field.  They report that a damping model could explain rims
in the younger remnants Cassiopeia A, Tycho, and Kepler, but not in SN
1006 or G347.3--0.5 (RX J1713.7-3946).  For the objects which satisfy
their conditions for magnetic damping, they deduce quite high
magnetic-field strengths of 200 -- 300 $\mu$G.

\section{Generalized Diffusion Model}

Here, we will consider the case of diffusion coefficients of the form
$D = \eta D_{B}(E_{h})(E/E_{h})^{\mu},$ where $D_{B}(E_{h})$ is the
Bohm diffusion coefficient at an arbitrary fiducial energy $E_{h}$ and a magnetic field $B_0$,
$\eta$ is a constant scaling factor taken in conjunction with
$D_{B}(E_{h})$ as a free parameter, and $\mu$ parameterizes the energy
dependence of $D$.  (So for Bohm diffusion, $\eta = 1$ and $\mu = 1.$) For $\mu<1$,
$E_{h}$ must be above the relevant energy range for X-ray emitting
electrons, so that $D$ remains greater than the minimum Bohm value at
all energies. On the other hand, for $\mu > 1$, this energy is the
lower threshold energy for this exotic type of diffusion to occur.  We
expect $\mu$ to be related to the power-law index $n$ of hydromagnetic
turbulence, $I(k) \propto k^{-n}$ where $I(k)$ is the wave power per
unit wavenumber.  Then $n = 5/3$ corresponds to a Kolmogorov spectrum,
and $n = 3/2$ to a Kraichnan spectrum.  In quasi-linear theory,
particles have a mean free path inversely proportional to the energy
density of MHD waves with wavelength comparable to the particle
gyroradius, resulting in $\mu = 2 - n$ (e.g., Reynolds 2004).  So
Kolmogorov turbulence predicts $\mu = 1/3$ and Kraichnan, $\mu = 1/2$.

In all of the ensuing discussion we will be concerning ourselves with
the consequences of these models observable in the X-ray filaments of
SN1006, which we will characterize by their FWHM, $x_{1/2}$, and its
energy dependence, again parametrized by $m_{\rm E}$.  This is
explicitly written as $x_{1/2}\propto E^{\rm m_{\rm E}}$.

For the case of $\mu=1$ the $E_h$ independent case, we adopt Rettig
$\&$ Pohl's solution to equation (12) for the electron spatial
distribution, assuming the injected spectrum to be an exponentially
cut off power law with index $s$, integrated over $n \equiv
E^\prime/E$ :
\begin{equation}
\begin{split}
f(x,E) = Q_{0} E^{-(s+1)} \int\limits^{\infty}_{1} \frac{n^{-s}}{\sqrt{\ln{(n)}}} \\
\times \exp\left[-\frac{nE}{E_{\rm cut}}-\frac{\left[l_{\rm ad}\left(1-\frac{1}{n}\right)-z(x)\right]^{2}}{4l_{\rm diff}^{2}\ln{(n)}}\right]dn,
\end{split}
\end{equation}
while for $\mu \ne$ 1, we adopt Lerche $\&$ Schlickeiser's (1980)
solution to equation (12),
\begin{equation}
\begin{split}
f(x,E) = \frac{Q_{0}}{\sqrt{(1-\mu)}} E^{-(s+1/2+\mu/2)} \int\limits^{1}_{0} \frac{n^{(s+\mu -2)/(1-\mu)}}{\sqrt{1-n}} \\
\times \exp\left[-\frac{n^{1/(1-\mu)}E}{E_{\rm cut}}-\frac{(1-\mu)\left[l_{\rm ad}\left(1-n^{1/(1-\mu)}\right)-z(x)\right]^{2}}{4l_{\rm diff}^{2}(1-n)}\right]dn,
\end{split}
\end{equation}
for $\mu<1$, and 
\begin{equation}
\begin{split}
f(x,E) = \frac{Q_{0}}{\sqrt{(1-\mu)}} E^{-(s+1/2+\mu/2)} \int\limits^{\infty}_{1} \frac{n^{(s+\mu -2)/(1-\mu)}}{\sqrt{n-1}} \\
\times \exp\left[-\frac{n^{1/(1-\mu)}E}{E_{\rm cut}}-\frac{(1-\mu)\left[l_{\rm ad}\left(1-n^{1/(1-\mu)}\right)-z(x)\right]^{2}}{4l_{\rm diff}^{2}(1-n)}\right]dn,
\end{split}
\end{equation}
for $\mu>1$.  Here $Q_{0}$ is a normalization constant that does not
factor into our calculations.  In this formalism, all the information
about the spatial dependence of the magnetic field is contained in the
function $z(x)$, defined as
\begin{equation}
z(x) = \frac{1}{B_{0}^{2}}\int\limits_{0}^{x}B(u)^{2}du.
\end{equation}  
where $B_0$ is the magnetic field immediately behind the shock, not the far upstream value.  It is presumably amplified from its initial value to the extent demanded by the data.

Furthermore, the cut-off energy, $E_{\rm cut}$, is found by equating loss
times and acceleration times, which gives, in the Bohm Limit (Rettig $\&$ Pohl 2012):
\begin{equation}
E_{\rm cut} = 8.3\ {\rm TeV}\ \left(\frac{B_{0}}{100 \mu{\rm
    G}}\right)^{-1/2} \left(\frac{v_{s}}{1000\ {\rm
    km\ s}^{-1}}\right).
\end{equation}
For arbitrary diffusion coefficients, it is a straightforward generalization to show:
\begin{equation}
E_{\rm cut} \propto \left(\frac{B_{0}}{100 \mu{\rm
    G}}\right)^{-\frac{1}{1+\mu}} \left(\frac{v_{s}}{1000\ {\rm
    km\ s}^{-1}}\right)^{\frac{2}{1+\mu}}\left(\frac{E_h}{\eta}\right)^{\frac{1}{1+\mu}}.
\end{equation}
These analytic solutions were derived under the assumption that
$B(x)^{2}D(x)$ is a constant with respect to $x$, the distance from the shock.  This condition is somewhat peculiar in that it
is only naturally satisfied if $B$ is constant, since the spatial dependence of $D(x)$ is contained in its dependence on $B$. For a field that evolves due to flux conservation, we expect this to be a reasonable
approximation within a thin rim. On the other hand, for a rapidly varying magnetic field (e.g. one that exponentially decays in space), this imposes a rapid variation in $D(x)$ with $x$, which may or may not be realistic.  However, we do not expect this to affect our results at the qualitative level, and some justification for this assumption can be found in Rettig $\&$ Pohl (2012).
 
Using this spectrum of electrons, intensity profiles are then obtained
by first evaluating the synchrotron emissivity
\begin{equation}
j_{\nu} = c_{3}B\int\limits^{\infty}_{0}G(y)f(x,E)dE
\end{equation}
with $y \equiv \nu/c_{1}E^{2}B$, and $G(y) \equiv y \int_y^\infty
K_{5/3}(z) dz$, in a slightly different notation from Pacholczyk
(1970); here $K_{5/3}(z)$ is a Bessel function of the second kind
with imaginary argument. Then, integrating along lines of sight:
\begin{equation}
I_{\nu}(r) = 2\int\limits^{\sqrt{r_{s}^{2}-r^{2}}}_{0}j_{\nu}\left(r_{s}-\sqrt{s^{2}+r^{2}}\right)ds.
\end{equation}

To characterize the size of the filaments, we use the FWHM of this
radial intensity, denoted $x_{1/2}$.  Then, we write $x_{1/2} \propto
E_{\gamma}^{m_{E}}$ to characterize the energy dependence of the FWHM
at each photon energy by
\begin{equation}
m_{E} = \frac{\log(x_{1/2}/x_{1/2}')}{\log(E_{\gamma}/E_{\gamma}')}.
\end{equation} 

In a brief aside, we note that one can get qualitative results for the
behavior of $m_E$ when $\mu \ne 1$ by noting that $l_{\rm diff} =
\sqrt{D\tau_{\rm synch}} \propto E^{(\mu-1)/2} \propto
\nu^{(\mu-1)/4}$ in the delta function approximation. Furthermore, we
can generalize equation (13) by using the new diffusion coefficient in
equation (6).  This gives
\begin{equation}
m_{E} = -\frac{1}{2}\left(\mu-\frac{4(\mu+1)D/(v_{\rm d}^{2}\tau)}{2+8D/(v_{\rm d}^{2}\tau)-2\sqrt{1+4D/(v_{\rm d}^{2}\tau)}}\right).
\end{equation}
Now at a photon energy of 2 keV, we can write $D$(2 keV) =
$\eta_{2}D_{B}(2 keV)$, and coupled with a modified equation (9), we
can solve uniquely for $\eta_{2}$ and the maximum field strength $B_{0}$ by measuring $w_{obs}$
and $m_{E}$ at 2 keV.  The results of using this approximate result
are shown in Table~\ref{tab:anal}.

\begin{deluxetable}{c c c c c c c}
\tablecaption{Best fit parameters for the Filaments in varying values of $\mu$ (Analytic Results)}
\tablewidth{0pt}
\tabletypesize{\scriptsize}
\tablenotetext{*}{Results of fitting equation (6) to the data using a Levenberg-Marquardt least-squares algorithm.  The stated uncertainties are estimated as 1$\sigma$. $B_{0}$ is in units of $\mu$G while $\eta_{2}$ is a dimensionless quantity representing the ratio of the fitted diffusion coefficient to the Bohm-limit diffusion coefficient at a photon energy of 2 keV. }
\tablehead{ &\multicolumn{2}{c}{Filament 1} & \multicolumn{2}{c}{Filament 2} & \multicolumn{2}{c}{Filament 3} }
	\startdata
$\mu$ &$\eta_{2}$ & $B_{0}$ &$\eta_{2}$ & $B_{0}$ & $\eta_{2}$ & $B_{0}$  \\ [.5ex]
0 & 7 $\pm$ 4  & 165 $\pm$ 21 & 0 $\pm$ 0.04 & 143 $\pm$ 52 & 0 $\pm$ 0.007 & 81 $\pm$ 3 \\
1/3 & 2.4 $\pm$ 0.9 & 130$\pm$ 8 & 0 $\pm$ 1.2 & 143 $\pm$ 39 & 0 $\pm$ 0.008 & 80.7 $\pm$ 0.9 \\
1/2 & 1.8 $\pm$ 0.6 & 123$\pm$ 7 & 0 $\pm$ 1.2 & 144 $\pm$ 35 & 0 $\pm$ 0.007 & 80.7 $\pm$ 1.2 \\
1 & 1.1$\pm$0.4& 113 $\pm$ 4 & 0 $\pm$ 1.1 & 145 $\pm$ 26 & 0 $\pm$ 0.018 & 80.7 $\pm$ 2 \\
1.5 & .8 $\pm$0.3& 108 $\pm$ 3 & 0 $\pm$ 1.0 & 145 $\pm$ 21 & 0 $\pm$ 0.03 & 80.7 $\pm$ 1  \\
2 & .7$\pm$0.3& 105 $\pm$ 3 & .2 $\pm$ 0.9 & 150 $\pm$ 21 & 0 $\pm$ 3$\times$10$^{-7}$ & 80.7 $\pm$ 1.1  \\
\hline
\hline 
\\
& &\multicolumn{2}{c}{Filament 4} && \multicolumn{2}{c}{Filament 5} \\
\hline
\\
$\mu$ &&$\eta_{2}$ & $B_{0}$ &&$\eta_{2}$ & $B_{0}$\\
0 && 0 $\pm$0.001 & 118 $\pm$ 1.1 && 0 $\pm$ 1.3$\times$ 10$^5$& 1000 $\pm$ 500\\
1/3 && 0 $\pm$ 2 $\times$ 10$^{-5}$ & 117.9 $\pm$ 0.8 && 0 $\pm$ 2 $\times$ 10$^4$ & 0 $\pm$ 60000\\
1/2 && 0 $\pm$ 0.0003 & 117.9 $\pm$ 1.4 && 0 $\pm$ 120  & 300 $\pm$ 300\\
1 &&  0 $\pm$ 0.016 & 117.9 $\pm$ 0.8 && 5 $\pm$ 5  & 160 $\pm$ 40\\
1.5 &&  0 $\pm$ 0.0012 & 117.9 $\pm$ 0.8 && 2.5 $\pm$ 1.7  & 140 $\pm$ 20\\
2 && 0 $\pm$ 1$\times$10$^{-8}$ & 117.9 $\pm$ 1.5 && 1.7 $\pm$ 0.9  & 125 $\pm$ 9\\
 
\enddata
\label{tab:anal}
\end{deluxetable}

When performing the full numerical calculation of FWHMs, we
distinguish between two parametrizations of the magnetic field, called
the ``loss-limited model'' and the ``magnetically damped model.''
That is, we can use the appropriate electron distribution (14), (15),
or (16) for both cases, varying the spatial dependence of $B$ to
select either loss-limited or magnetically damped situations.  We will
initially neglect the cut-off in the injected electron spectrum in
order to more clearly highlight the energy dependence of each model;
we include the effects of cut-offs in Section 3.4.

\subsection{Loss-Limited Model}

In the loss-limited model, we assume the magnetic field is spatially
uniform, a good approximation if we expect it to evolve by flux
conservation in the narrow region behind the shock.  Then the function
$z(x)$ as defined in Equation (17) reduces to just $x$.  With only two free
parameters, the scaling factor for the diffusion coefficient at 2 keV,
$\eta_{2}$ and the maximum field strength $B_{0}$, we can fit the
observed filaments and their energy dependence uniquely for any value
of $\mu$.  For the case of Bohm diffusion ($\mu=1$), the simple
estimate of section 1.1.1 resulted in an equation (13) for $m_{E}$
which is independent of $B_{0}$. This behavior is preserved in the
full calculation, and even for values of $\mu$ differing from 1 the
energy dependence only weakly depends on the magnetic field strength.

\subsubsection{Energy Dependence of the Loss-Limited Model}

\begin{figure}[H]
\centering
\includegraphics[scale=.4]{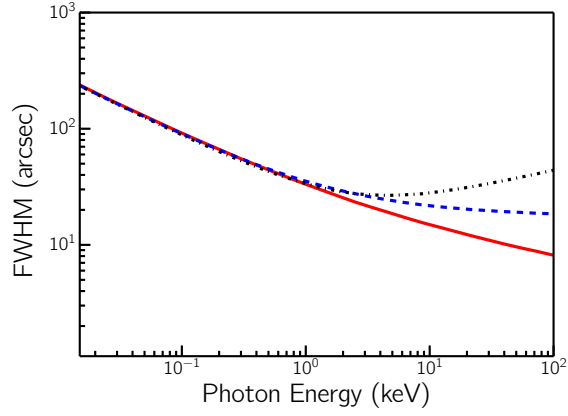}
\caption{Energy dependence of filament widths for 
different diffusion coefficients, for pure power-law electron
spectra without a cut-off. Solid line is Kolmogorov-like ($D \propto E^{1/3}$), dashed  line is Bohm-like ($D \propto E$), and dot-dashed line is for $\mu = 2$ ($D \propto E^2$).}
\label{meplot}
\end{figure}

\begin{figure}[H]
\centering
\includegraphics[scale=.4]{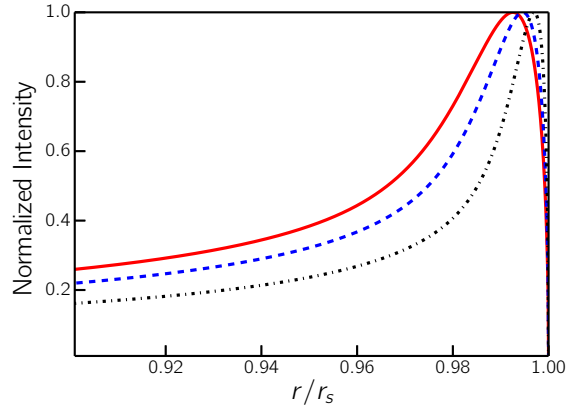}
\caption{Calculated profiles in the loss-limited model for $B_0 = 100 \mu G$, $\mu = 1/3$.  The solid line represents a photon energy of 1 keV, the dashed line represents a photon energy of 2 keV, and the dot-dashed line represents a photon energy of 8 keV.}
\label{LossProfile}
\end{figure}

\begin{figure}[H]
\centering
\includegraphics[height=2.5in]{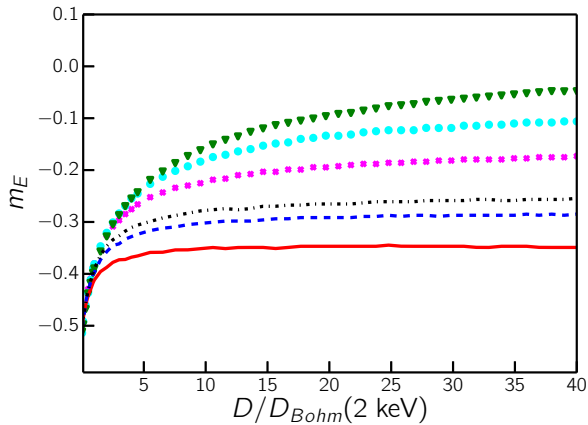}
\caption{Dependence of the parameter $m_{\rm E}$ on the magnitude of the diffusion coefficient, $D$, measured in units of the Bohm value at a photon energy of 2 keV.  Recall that $m_{\rm E}$ is defined such that the FWHM of the rim is $\propto E^{m_{\rm E}}$.  The lines are, from bottom to top: $\mu = 0, 1/3, 1/2, 1$ (Bohm), $1.5,$ and $2$.  The calculations were done with $B_0 = 100 \mu G$. For small values of $D$ we find that all models converge to just below $m_{\rm E} = -0.5$, while for larger values of $D$ clear limits can be placed on the range of $m_{\rm E}$, an observable quantity, for each value of $\mu$ (where $D \propto E^{\mu}$).}
\label{Etagraph}
\end{figure}

In the loss limited model, we still see the same general behavior of
the FWHM with energy as the calculation of Section 1.2.  There is a
clear transition between energies where advection is dominant to those
where diffusion is dominant, with $m_{E}$ dropping from --1/2 to
($\mu-1)/4$. A plot of this behavior for several values of $\mu$ is
shown in Figure~\ref{meplot} and an example of calculated profiles is
shown in Figure~\ref{LossProfile}. A crucial point to make is that the
magnitude of the diffusion coefficient is by far the most important
factor in determining $m_{\rm E}$.  For Bohm-type diffusion
(i.e. $\mu=1$) this mapping is 1-1, while for $\mu \neq 1$ there
exists only a weak dependence of $m_{\rm E}$ on the magnitude of the
post-shock magnetic field. Thus, the observation of $m_{\rm E}$ is a
direct probe of the properties of the diffusion coefficient, including
both its magnitude and behavior with energy, as can be clearly seen in
Figure~\ref{Etagraph}.

\subsection{Magnetically Damped Model}

In this model, we assume that the magnetic field amplification
decays exponentially behind the shock, in the form of $B(x) = B_{min}
+ \left(B_{0}+B_{min}\right)e^{-x/a_{b}}$, with $B_{min}$ taken to be
$5 \mu G$ (conservatively taken to be slightly higher with the above quoted value of 3 $\mu$G in the ISM). For this damped form of the magnetic field,
\begin{equation}
z(x) = \left(\frac{B_{min}}{B_{0}}\right)^{2}x+2a_{b}\frac{B_{min}(B_{0}-B_{min})}{B_{0}^{2}}(1-e^{-x/a_{b}})+\frac{a_{b}}{2}\left(\frac{B_{0}-B_{min}}{B_{0}}\right)^{2}(1-e^{-2x/a_{b}}).
\end{equation}
In order to both produce the observed filament profiles and to be
distinguishable from the loss-limited model,
$B_{0}$ must be roughly at least four times $B_{min}$.  There are 
three free parameters, $\eta$, $B_{0}$, and $a_{b}$, so with
only two observational constraints the fits are not unique.

\subsubsection{Energy Dependence in the Magnetically Damped Model}
\begin{figure}[H]
\centering
\includegraphics[scale=.3]{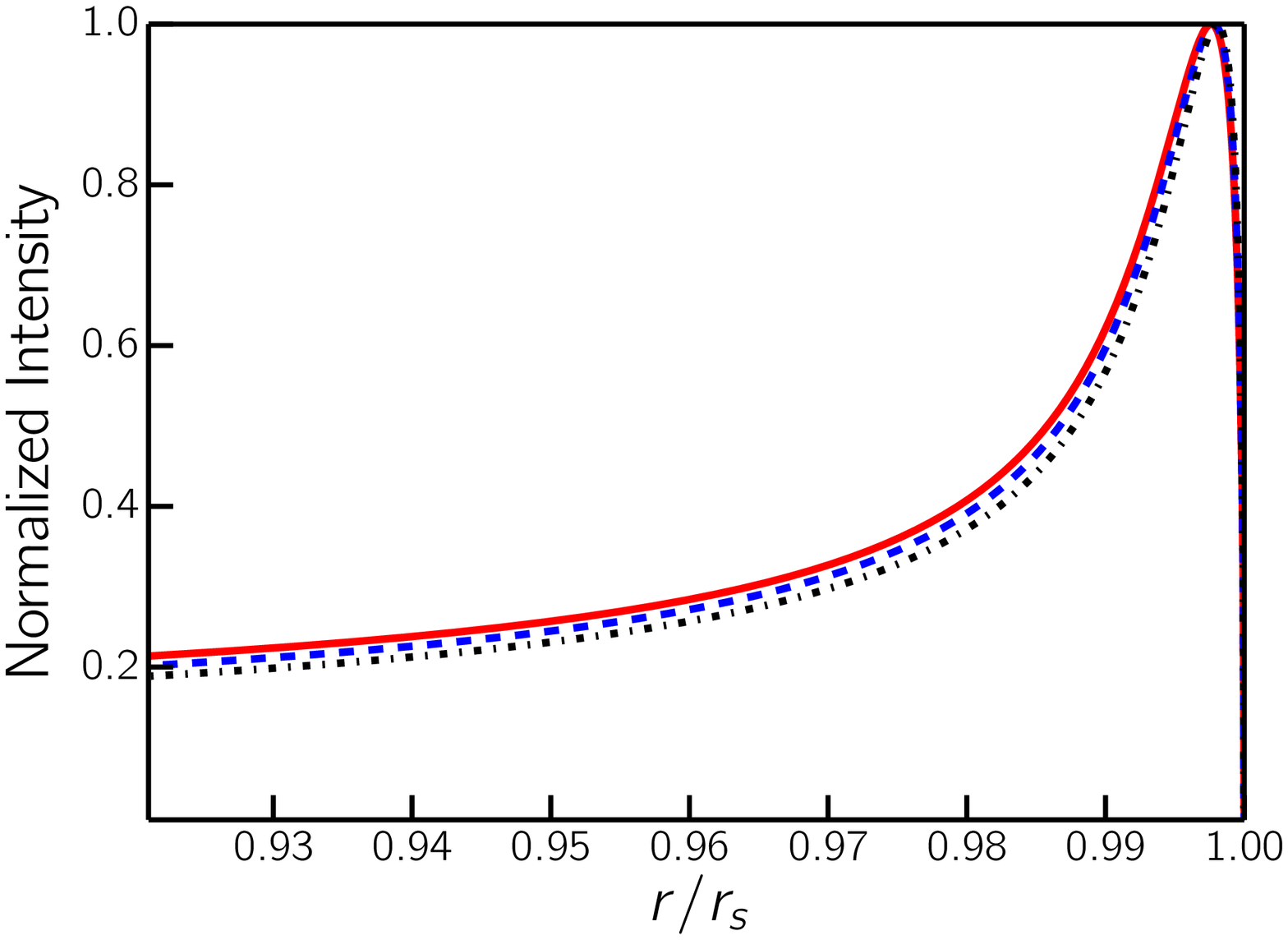}
\includegraphics[scale=.3]{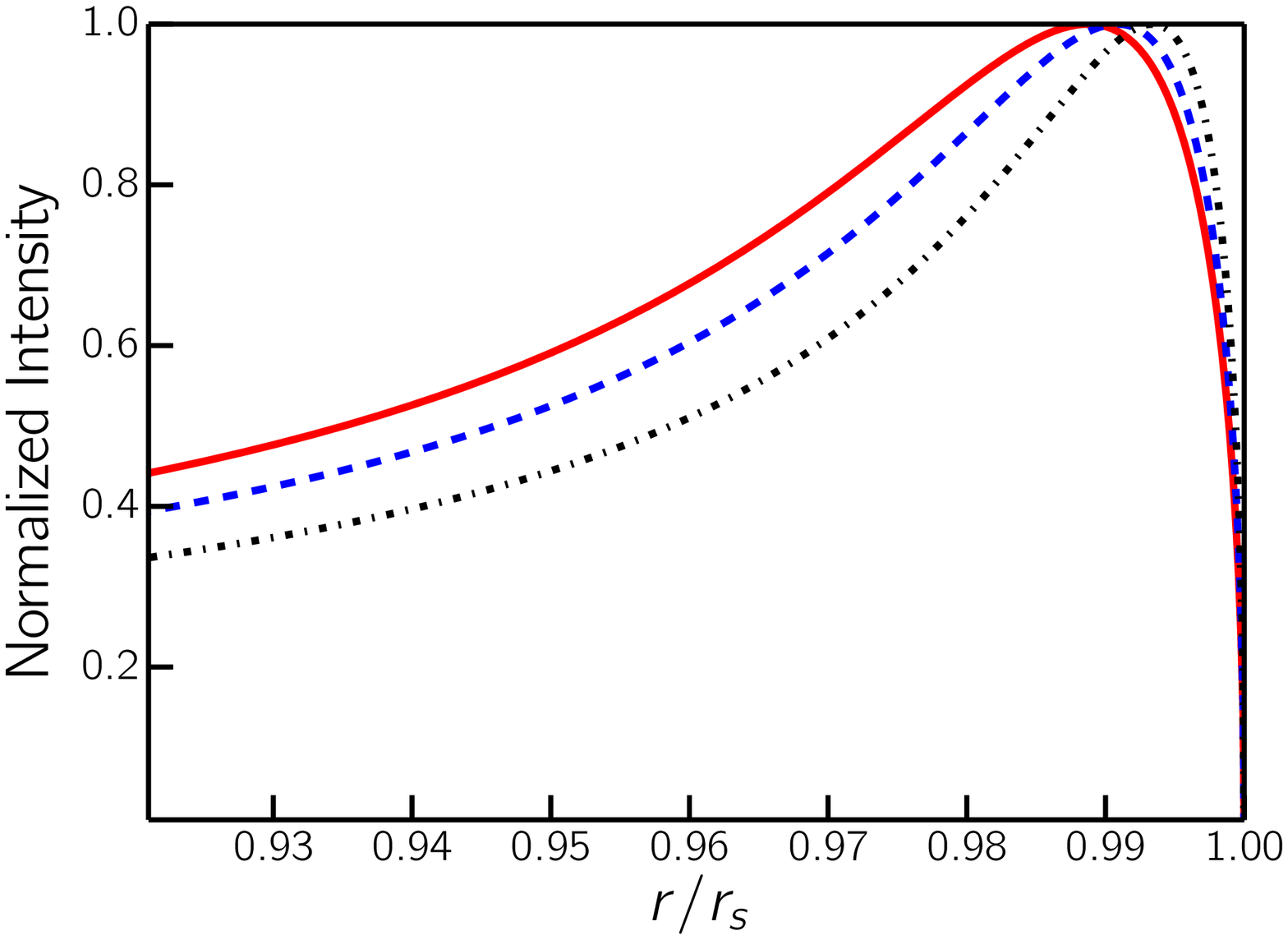}
\caption{Calculated profiles in the magnetically damped model for $B_0 = 70 \mu G$, $a_b = .005r_s$ (left) and $a_b = .05r_s$ (right). The solid line represents a photon energy of 1 keV, the dashed line represents a photon energy of 2 keV, and the dot-dashed line represents a photon energy of 8 keV.}
\label{DampProfile}
\end{figure}

Again, neglecting the energy cut-off for a moment, we see key features
develop in the energy dependence of the FWHM.  At low photon energies, where
losses are negligible over the small region $a_{b}$, rim sizes are  
energy-independent.  Radial profiles at these energies reflect
the spatial dependence of the magnetic field, and so, if this model is correct, we would expect thin 
filaments to be observed in high resolution radio images. At higher energies, the maximum value that $m_{E}$
reaches is determined by the competition between $a_{b}$, $l_{\rm diff}$,
and $l_{\rm ad}$.  This can be roughly expressed by the equation
\begin{equation}
l_{\rm eff} = min\left[max\left(l_{\rm ad},l_{\rm diff}\right), a_{b}\right].
\end{equation}
Thus there are three possibilities.  If $l_{\rm diff}$ is small enough,
then as energy increases, synchrotron losses will ``catch up" to
$a_{b}$ and there will be a clear transition between loss-limited rims
and magnetically limited rims.  If $l_{\rm diff}$ is large enough, the
rims will be damping limited at all photon energies. Finally, if
$a_{b}$ is large enough, the rims will be loss-limited at all photon
energies. It is worth noting that low-energy (i.e., radio) thin
synchrotron filaments would be a clear signature of field damping. Examples of how the calculated profiles can vary with energy in 'strong' and 'weak' damping are plotted in Figure~\ref{DampProfile}.

\subsection{Electron Cut-Off Energy and the Energy 
Dependence of Rim Widths}

The fact that the injected spectrum of electrons has 
a cut-off above some energy $E_{\rm cut}$ has an
impact on the energy dependence of the FWHMs of the observed intensity
profiles.  This is most easily seen in the $\delta$-function
approximation to the synchrotron emissivity for the case of a constant
magnetic field $B_{0}$ in the absence of a cut-off in the electron distribution. Here
$z(x)$ is just $x$, $E = \sqrt{\nu/(c_m B_0)} \equiv E_{\nu,0}$ and, using
\begin{equation}
f(E) = K(E_{0}(E))^{-s}\frac{dE_0}{dE}
\end{equation}
and
\begin{equation}
E_0 = \frac{E}{1-EbB_0^2t} = \frac{Ev}{v-EbB_0^2x} \equiv \frac{E_{\nu,0}}{1-x/l_{\rm ad}} 
\end{equation}
from Reynolds (2009), we get for the spatial dependence of the emissivity (recalling that $j_\nu \propto \sqrt{\nu B}f(\nu,x)$ in this case)
\begin{equation}
j_{\nu} = C_{j} \left(1-\frac{x}{l_{\rm ad}}\right)^{s-2}e^{\frac{-E_{\nu,0}}{E_{\rm cut}\left(1-\frac{x}{l_{\rm ad}}\right)}}.
\end{equation}

\begin{figure}[h]
\includegraphics[scale = .4]{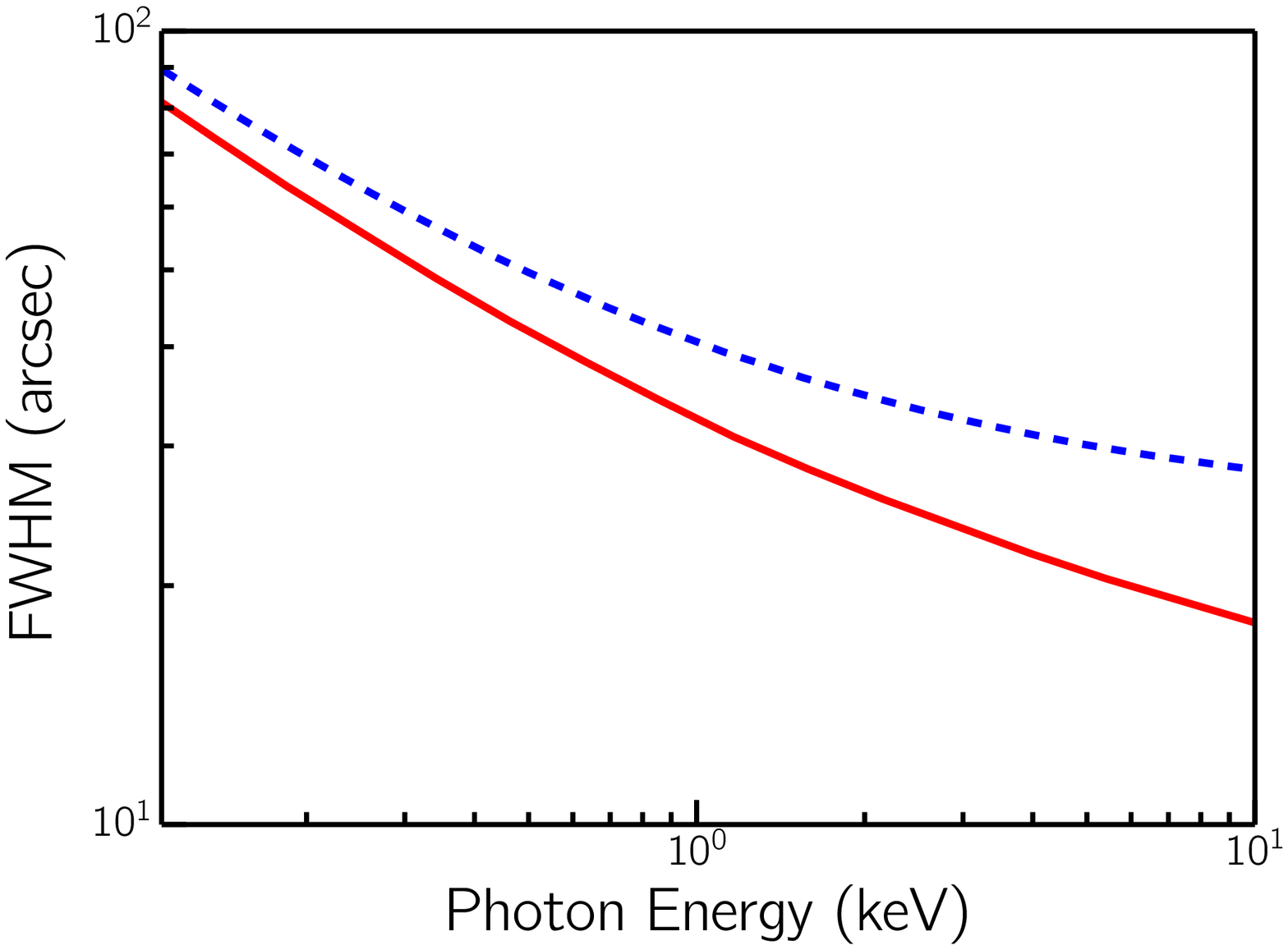}
\includegraphics[scale=.4]{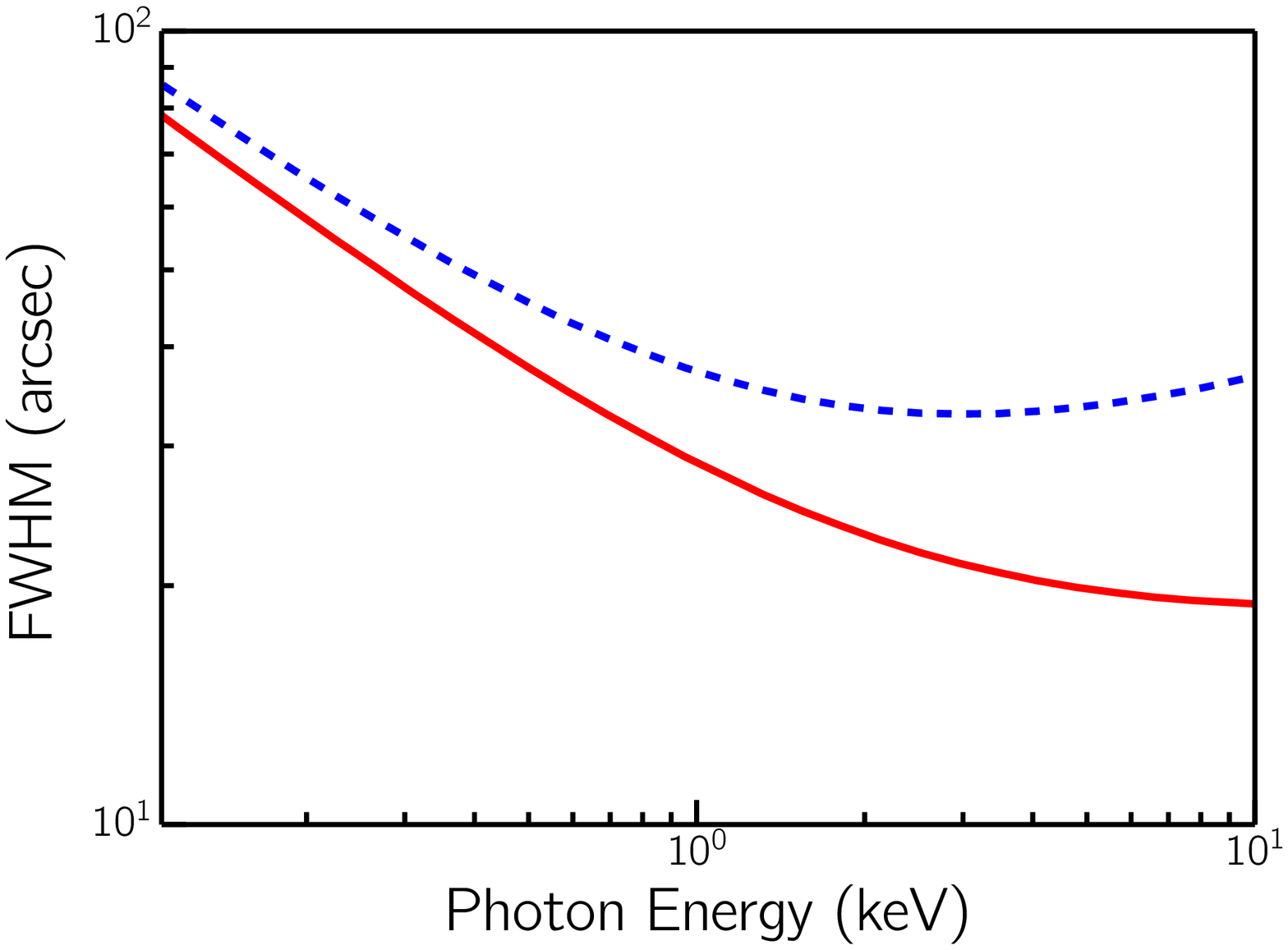}
\caption{Demonstration of the effect of including a cut-off in the injected electron spectrum on the energy dependence of the filament widths for $\eta_2$ = 2.6, $B_{0} = 100 \mu G$, and for $\mu =1$(left) and $\mu = 2$(right). Solid lines represent calculations with a cut-off in the electron spectrum and dashed lines represent calculations without a cut-off in the electron spectrum.}
\label{cutoff}
\end{figure}

When $E \gapprox E_{\rm cut}$, the exponential term dominates the spatial
behavior, resulting in an emissivity that will decay to half its peak
at a distance $x_{1/2}$ given by
\begin{equation}
x_{1/2} = \frac{l_{\rm ad}}{1+\frac{E_{\nu,0}}{E_{\rm cut}\ln(2)}}.
\end{equation}
with an energy index of
\begin{equation}
m_{E} = \frac{\partial \log{(x_{1/2})}}{\partial \log{(\nu)}} = -\frac{1}{2}\left(1+\frac{\nu^{1/2}}{E_{\rm cut}\sqrt{c_{m}B_{0}}+\nu^{1/2}}\right) =- \frac{1}{2}\left(1+\frac{E_{\gamma}^{1/2}}{E_{\rm rolloff}^{1/2}+E_{\gamma}^{1/2}}\right)
\end{equation}
So near the rolloff photon energy $E_{\rm rolloff} \equiv h\nu_{\rm m}(E_{\rm cut})$, 
$|m_{E}|$ is higher that the value of 
$1/2$ expected from pure advection.  In the full numerical
calculation, this arises as a shift in the expected $m_{E}$ by some
negative constant above some energy, even when diffusion is the
dominant method of transport (see Figure~\ref{cutoff}).

\subsection{Shifts in Intensity Peak Location}

In Figures~\ref{LossProfile} and \ref{DampProfile} we see that in the presence of filament widths that shrink with energy there is an associated outward motion of the location of peak emission.  Regardless of the underlying mechanism responsible for reducing the widths, that same mechanism explains this peak shift as it focuses the emission to increasingly narrow regions at higher energies. Thus this effect is a model-independent prediction.  

However, determining the location of the peak emission from observations is a task much more uncertain than determining the FWHM of radial profiles.  Furthermore, the predicted shift in location can be quite small ($\lapprox 1'$) for the best-fit parameters found in section 4. The combination of these two limitations leads us to ignore this effect in the rest of our analysis.

\subsection{Comparison to the $\delta$-Function Approximation}

\begin{figure}[h]
\centering
\includegraphics[scale=.4]{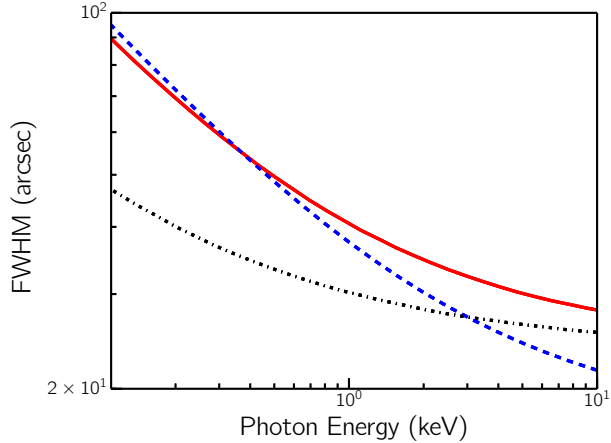}
\caption{Comparison of the energy dependence of the FHWM for Bohm diffusion at $B_0 = 100 \mu G$ and $\eta$=2.6 in different possible approximations without an electron cut-off. Similar differences were seen for other input parameters. Solid line represents a full convolution without a cut-off in the electron spectrum, dashed line represents the $\delta$-function approximation used with the catastrophic dump convection-diffusion equation, and dot-dashed line represents the $\delta$-function approximation used with the continuous energy loss convection-diffusion equation.}
\label{approxs}
\end{figure}

We first considered using the delta function approximation of the
emissivity, namely
\begin{equation}
j_\nu \propto \sqrt{B\nu}f\left(x, E = \sqrt{\frac{\nu}{c_{m}B}}\right).
\end{equation}
However, when compared with the full convolution of the particle
distribution with the single electron emissivity, the results for the
FWHMs disagree considerably (see Figure~\ref{approxs}).  What is worse, the
difference is dependent on the electron energy so that a simple
constant correction factor could not be employed.  The combination of
the $\delta$-function approximation with the catastrophic dump form of
the convection-diffusion equation seems to provide a much better
approximation, but it does not  account for the
cut-off in the injected spectrum of electrons. Thus, we were compelled
to use the full synchrotron emissivity in our numerical calculations
coupled with the integral solution to the continuous energy loss
convection-diffusion equation.

\section{Results}
In this section we first summarize our observational methodology in measuring the filament widths and spectra. Then we detail our fitting procedure for applying our model to the data and describe our findings.

\begin{figure}[H]
\centering
\includegraphics[scale=.5]{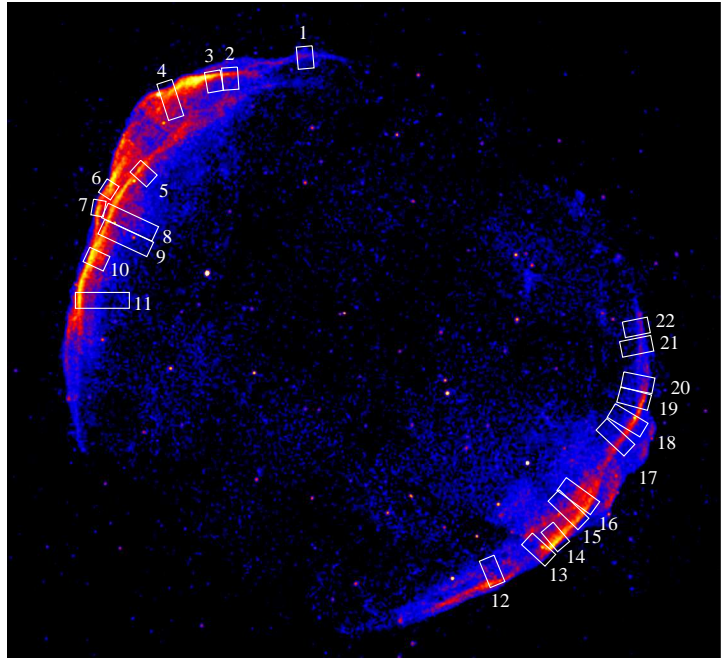}
\caption{\emph{Chandra} image at 2-7 keV showing the regions where radial profiles were extracted. Filament 1:Regions 1-4 and 6; Filament 2: Regions 5, 7, and 9-11; Filament 3: Regions 12-16; Filament 4: Regions 17-22; Filament 5: Regions 6 and 8}
\label{regions}
\end{figure}

To extract radial profiles of the NE and SW limbs of SN~1006, we use
six {\it Chandra} observations, the parameters of which are summarized
in Table~\ref{tab:obs}.  The observations of the SW and some of the NE
were performed as part of a {\it Chandra} Large Program (Winkler et
al. 2014).  These new observations provide the first high quality
image of the SW quadrant, comparable in quality with previous images
of the NE. We reprocessed the level-1 event files with CIAO ver.\ 4.4
and CALDB ver.4.5.1.  After correcting for vignetting effects and
exposure times for all of the data sets, we extract radial profiles in
three energy bands: 0.7--1\,keV, 1--2\,keV, and 2--7\,keV from 22
regions shown in Figure~\ref{regions}.  Each profile is binned by
1$^{\prime\prime}$.  When combining the NE profiles from different
epochs, we take into account the expansion of the remnant by 4"
according to the literature (Katsuda et al.\ 2009; Winkler et
al.\ 2014). Region 8 was excluded from this analysis because in the
lowest energy bin there was spatial overlap between two filaments.

\begin{deluxetable}{lccccccc}
\tabletypesize{\tiny}
\tablecaption{{\it Chandra} observations of SN~1006}
\tablewidth{0pt}
\tablehead{
\colhead{ObsID}&\colhead{Array}&\colhead{R.A. (J2000)}&\colhead{Decl. (J2000)}&\colhead{Roll}&\colhead{Obs.\ Date}&\colhead{Exposure (ks)}&\colhead{PI}}
\startdata
732 & ACIS-S & 15:03:51.7 & -41:51:16 & 280.2$^\circ$ & 2000 Jul 10 & 55.3 & K.S. Long \\
9107 & ACIS-S & 15:03:51.5 & -41:51:19 & 280.4$^\circ$ & 2008 Jun 24 & 68.9 & R. Petre\\
13738 & ACIS-I & 15:01:43.7 & -41:57:55 & 25.3$^\circ$ & 2012 Apr 23 & 73.5 & P.F. Winkler\\
13739 & ACIS-I & 15:02:14.9 & -42:06:49 & 9.1$^\circ$ & 2012 May 4 & 100.1 & P.F. Winkler\\
13743 & ACIS-I & 15:03:01.8 & -41:43:05 & 19.9$^\circ$ & 2012 Apr 25 & 92.6 & P.F. Winkler\\
14424 & ACIS-I & 15:01:43.7 & -41:57:55 & 253.1$^\circ$ & 2012 Apr 27 & 25.4 & P.F. Winkler\\
\enddata

\label{tab:obs}
\end{deluxetable}

\subsection{Profile Modeling}
\begin{figure}[H]
\centering
\includegraphics[angle = 0, width=\textwidth, trim=.5cm 0cm 0cm 0cm, clip=true]{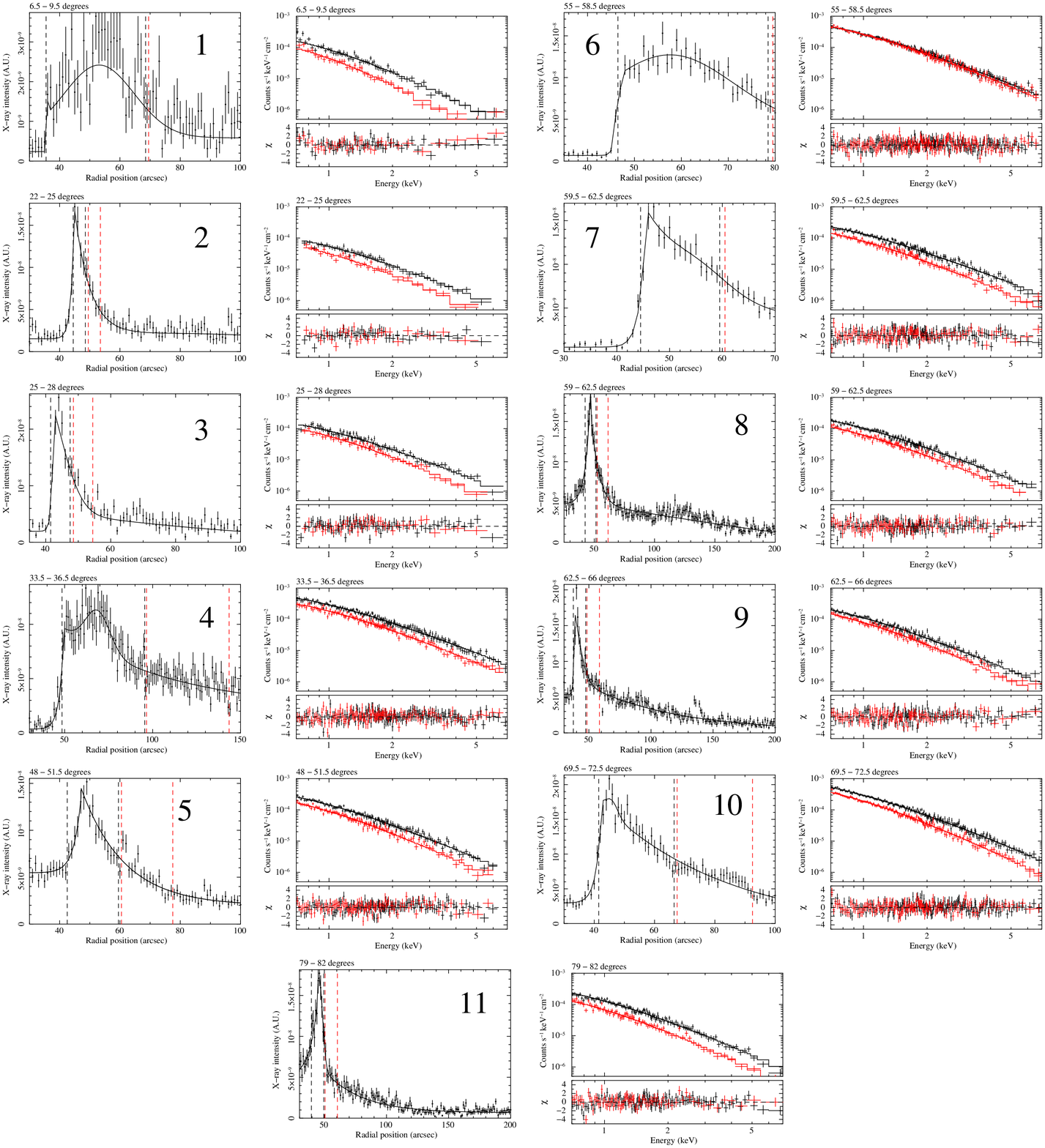}  
\caption{Left: Filament radial intensity profiles at 2-7 keV. The dashed, black vertical lines occur at a length of one FWHM on each side of the emission's peak, while the dashed, red vertical lines enclose the region extending towards the center of the remnant that starts at the edge of the black region to a distance of 2$\times$FWHM away from the peak.  Right: Energy spectra of the filaments separated into the same two regions}
\label{profiles12}
\end{figure}

\begin{figure}[H]
\centering
\includegraphics[angle = 0,width=\textwidth,trim=.5cm 0cm 0cm 0cm, clip=true]{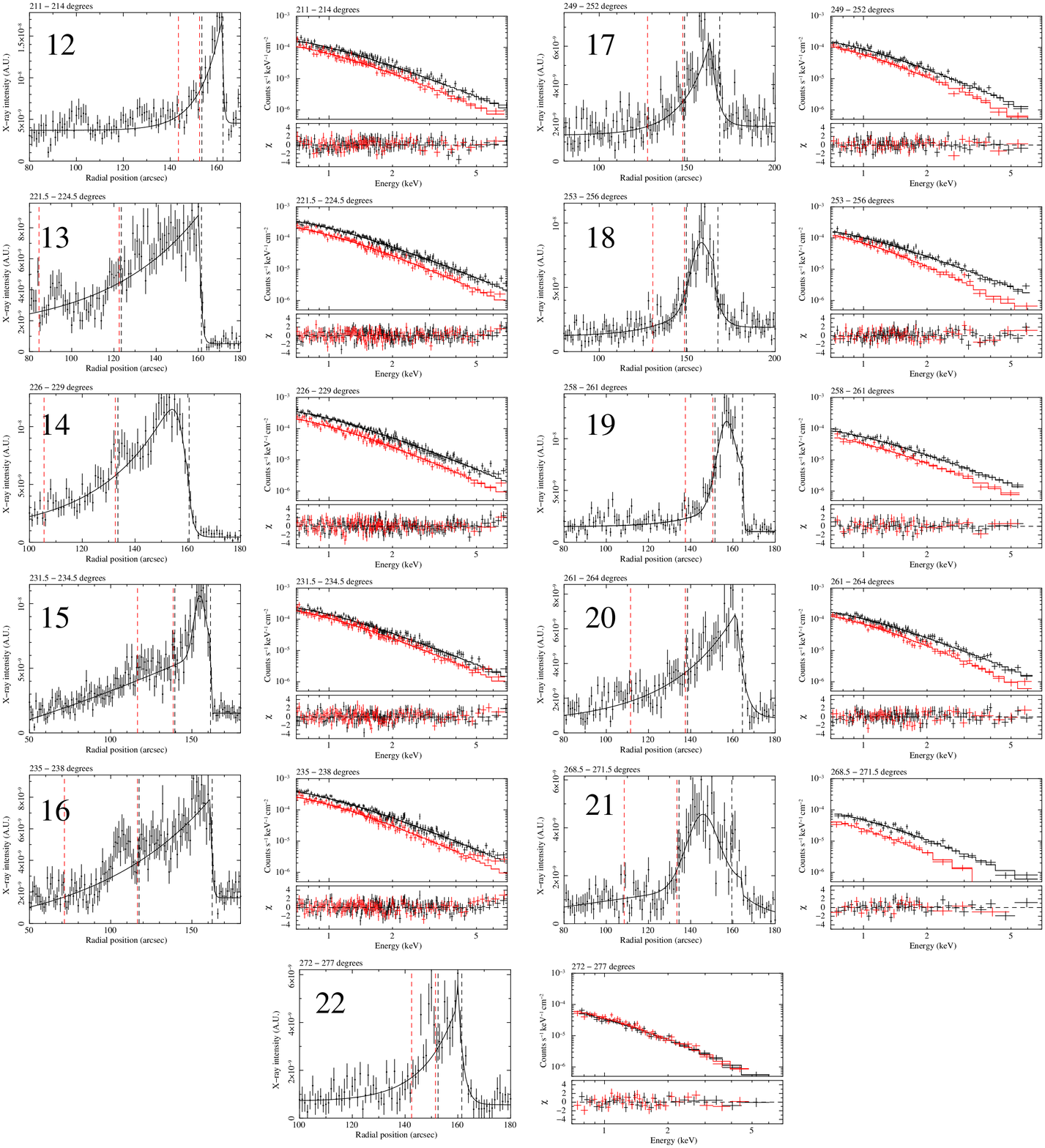}
\caption{See Figure \ref{profiles12} caption}
\label{profiles34}
\end{figure}

To estimate rim widths, we fit each profile with an empirical model defined as,
\begin{equation}
h(x) = \left \{
\begin{array}{l}
A_u~{\rm exp}\left(\frac{x_0 - x}{w_u}\right) + C_u~~~~~~{\rm (upstream)}\\
A_d~{\rm exp}\left(\frac{x - x_0}{w_d}\right) + B~{\rm exp}\left(\frac{-(x - x_1)^2}{2\pi\sigma^2}\right) + C_d~~~~~~{\rm (downstream)}
\end{array}
\right.
\end{equation}
where $A_u$, $x_0$, $w_u$, $C_u$, $A_d$, $w_d$, $B$, $x_1$, $\sigma$,
and $C_d$ are all free parameters.  We note that either $x_0$ or $x_1$ can correspond to the peak of the X-ray profile, and that $C_u$ represents
the background level.  The best-fit models are plotted as solid lines
in Figures~\ref{profiles12} and \ref{profiles34}.  Based on the
best-fit model, we calculate a full width at half maximum (FWHM) for
each profile.  The model accounts for plateaus of emission upstream and
downstream of the peak; the Gaussian component describes possible downstream
features due to projection effects.  Since our primary interest is in
the energy-dependence of widths, the most important consideration is
the consistency of a filament model among the three energy bins.  
To estimate the uncertainties of FWHMs, the best-fit profiles are
artificially re-scaled (stretched or shrunk) along the $x$-axis, so
that a new $x$-position of the model profile ($x'$) becomes $x\left(1
+ \xi\times\frac{x - x_0}{200^{\prime\prime} - x_0}\right)$, where $x$
is the original $x$-position of the model profile and $\xi$ is a
variable stretch factor.  For various $\xi$-values, $\chi^2$ values
between the re-scaled model profile and the data are calculated,
resulting in statistical uncertainties on $a$ (and FWHM).

The best-fit FWHMs and their statistical uncertainties (ranges
corresponding to $\Delta \chi^2 = 2.7$) are listed in
Tables~\ref{tab:FWHM1}, \ref{tab:FWHM2}, and \ref{tab:FWHM3}.  The results are
categorized into four groups that appear to be along the same
filaments. Also listed in Tables~\ref{tab:FWHM1},
\ref{tab:FWHM2}, and \ref{tab:FWHM3} is the average value of $m_{E} \equiv
\log($FWHM/FWHM$^\prime$)/$\log(\nu/\nu')=$ either $\log($FWHM/FWHM$^\prime$)/$\log(2/1)$ or $\log($FWHM/FWHM$^\prime$)/$\log(1/0.7)$
for our purposes, taking the FWHM for each energy range as that of the
lower limit and using the lower adjacent energy bin for the primed
variables. This quantity characterizes the energy dependence of the
FWHMs by writing them as $\propto$ $E^{\rm m_E}$. To get the
uncertainties on the calculation of $m_{\rm E}$ and the averages, we
took the uncertainty on each data point as approximately symmetric
with $\sigma = (\sigma_{+}+\sigma_{-})/2$.

\begin{figure}[H]
\centering
\includegraphics[scale=.4]{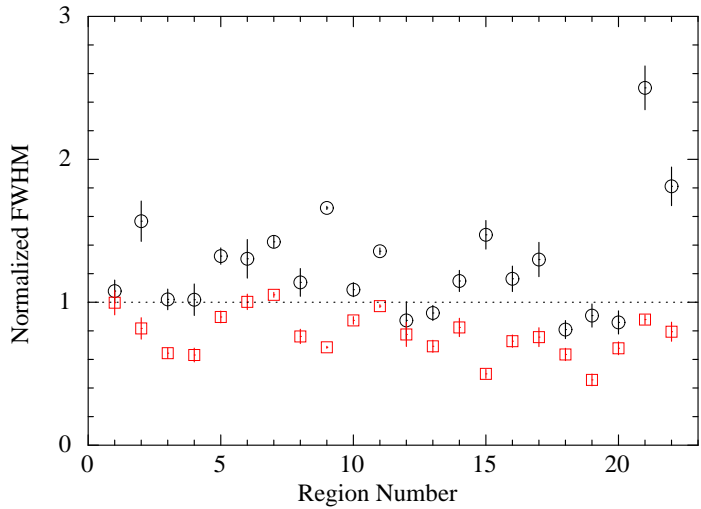}

\caption{Observed energy dependence of the FHWMs of SN1006, plotted vs. region number and normalized to the middle (1--2 keV) energy band. Circles represent 0.7--1 keV while squares represent 2--7 keV}
\label{widthdata}
\end{figure}

To check the energy-dependence of rim widths, we extract two X-ray
spectra from each region: one is taken from a filament region
(covering from a shock front to a FWHM position downstream) and the
other is taken from a plateau region next to the filament region up to
a 2$\times$FWHM position downstream.  These spectra together with the
best-fit models ({\tt srcut} in XSPEC: Reynolds 1998) are presented in
Figures~\ref{profiles12} and~\ref{profiles34}, where black and red are
responsible for the filament and the plateau regions, respectively.
In some regions (e.g., region \#3), spectral softening downstream is
clearly seen.  This is consistent with the fact that the higher the
energy band, the narrower the rim widths become, as shown in Figure \ref{widthdata}.

\begin{deluxetable}{c c c c c c c c}
\tabletypesize{\tiny}
\tablecaption{Measured Filament FWHM (arcseconds) vs. Energy Band - Northeast Limb}
\tablewidth{0pt}
\tablehead{\multicolumn{4}{c}{Filament 1} & \multicolumn{4}{c}{Filament 2}}
\tablenotetext{*}{ When calculating the uncertainties on the average FWHM and the
average $m_{E}$, the uncertainties on each individual FWHM were
treated as symmetric with uncertainty $(\sigma_{+}+\sigma_{-})/2$. The average $m_{\rm E}$ is defined as the $m_{\rm E}$ calculated from the average FWHMs. }
\startdata
Region &0.7-1 keV & 1-2 keV &2-7 keV  & Region & 0.7-1 keV & 1-2 keV & 2-7 keV\\ [.5ex]
1 & $36^{+2.3}_{-2.1}$  & $33.4^{+1.5}_{-1.0}$ &  $33.3^{+2.3}_{-2.8}$ & 5 & $25.8^{+.9}_{-1}$ & $19.5^{+0.3}_{-0.6}$ & $17.5^{+0.7}_{-0.7}$ \\
2 & $9.4^{+0.8}_{-0.7}$ & $6.0^{+0.2}_{-0.3}$ & $4.9^{+0.5}_{-0.3}$ & 7 & $13.8^{+0.4}_{-0.3}$ & $9.7^{+0.1}_{-0.2}$ & $10.2^{+0.1}_{-0}$\\
3 & $10.3^{+0.6}_{-0.6}$ & $10.1^{+0.4}_{-0.4}$& $6.5^{+0.3}_{-0.2}$ & 9 & $26.9^{+0.3}_{-0.6}$ & $16.2^{+0.1}_{-0}$ & $11.1^{+0.1}_{-0.1}$\\
4 & $78.1^{+7.3}_{-6.8}$ & $76.7^{+4.6}_{-4.6}$& $48.4^{+2.4}_{-2.2}$ & 10 & $33.4^{+1.5}_{-1.3}$ & $30.7^{+0.5}_{-0.5}$ & $26.8^{+0.7}_{-0.6}$\\
6 & $43.7^{+5.5}_{-3.3}$ & $33.5^{+0.8}_{-0.9}$ & $33.6^{+1.6}_{-1.6}$ & 11 & $15.2^{+0.3}_{-0.2}$ & $11.2^{+0.1}_{0}$ & $10.9^{+0.9}_{-0.7}$\\

Average& $36 \pm 1.7$  & 32 $\pm$ 1.0 & 25 $\pm$ 1.7 &Average& 23.0 $\pm$ 0.4 & 17.5 $\pm$ 0.14 & 15.3 $\pm$ 0.6 \\
Average $m_{E}$ &  & -0.30 $\pm$ 0.16 & -0.3 $\pm$ 0.11 & Average $m_{E}$ & & -0.78 $\pm$ 0.05 & -0.19 $\pm$ 0.05\\

\enddata
\label{tab:FWHM1}
\end{deluxetable}

\begin{deluxetable}{c c c c c c c c}
\tabletypesize{\tiny}
\tablecaption{Measured Filament FWHM (arcseconds) vs. Energy Band - Southwest Limb}
\tablewidth{0pt}
\tablehead{\multicolumn{4}{c}{Filament 3} & \multicolumn{4}{c}{Filament 4}}
\tablenotetext{*}{See note on Table \ref{tab:FWHM1}}
\startdata
Region &0.7-1 keV & 1-2 keV &2-7 keV & Region &0.7-1 keV & 1-2 keV &2-7 keV \\ [.5ex]
12 & $12.4^{+1.7}_{-1.6}$  & $14.2^{+1.0}_{-1.0}$ &  $11.0^{+0.9}_{-0.9}$ & 17 &  $35.2^{+2.8}_{-3}$ & $27.1^{+1.2}_{-1.1}$ & $20.5^{+1.6}_{-1.5}$\\
13 & $50.8^{+2.6}_{-1.9}$  & $54.9^{+2.3}_{-1.7}$ &  $38.0^{+2.9}_{-0.8}$ & 18 & $24.2^{+1.5}_{-2.0}$ & $29.9^{+0.9}_{-0.9}$ & $19.0^{+1.2}_{-1.1}$\\
14 & $38.6^{+2.2}_{-1.9}$  & $33.6^{+1.3}_{-1.1}$ &  $27.7^{+3.2}_{-0.6}$ & 19 & $13.7^{+1.1}_{-1.1}$ & $15.1^{+0.6}_{-0.6}$ & $6.9^{+0.6}_{-0.5}$\\
15 & $69.9^{+3.8}_{-4.5}$  & $47.5^{+1.1}_{-1.9}$ &  $23.7^{+1.5}_{-1.0}$ & 20  & $34.2^{+3.0}_{-2.9}$ & $39.8^{+1.5}_{-1.6}$ & $27.0^{+0.1.6}_{-1.3}$\\
16 & $74.0^{+5.2}_{-5.1}$  & $63.6^{+2.1}_{-2.0}$ &  $46.3^{+2.3}_{-2.3}$ & 21 & $35.0^{+1.7}_{-2.1}$ & $14.0^{+0.9}_{-0.1}$ & $12.3^{+0.1}_{-0.5}$\\
 & & & & 22  & $31.7^{+2.2}_{-1.9}$ & $17.5^{+0.5}_{-0.8}$ & $13.9^{+0.9}_{-1.2}$\\

Average & 49 $\pm$ 1.5 & 42.8 $\pm$ 0.7 & 29.3 $\pm$ 0.8 & Average & 29.0 $\pm$ 0.9 & 23.9 $\pm$ 0.4 & 16.6 $\pm$ 0.5 \\
Average $m_{E}$ &  & -0.4 $\pm$ 0.10 & -0.54 $\pm$ 0.04 & Average $m_{E}$ & & -0.5 $\pm$ 0.10 & -0.53 $\pm$ 0.05\\

\enddata
\label{tab:FWHM2}
\end{deluxetable}

\begin{deluxetable}{c c c c }
\tabletypesize{\tiny}
\tablecaption{Measured Filament FWHM (arcseconds) vs. Energy Band - Southwest Limb}
\tablewidth{0pt}
\tablehead{\multicolumn{4}{c}{Filament 5} }
\tablenotetext{*}{See note on Table \ref{tab:FWHM1}}
\startdata
Region &0.7-1 keV & 1-2 keV &2-7 keV \\ [.5ex]
8 & $23.8^{+2.0}_{-1.5}$  & $20.9^{+1.0}_{-0.8}$ &  $15.9^{+0.8}_{-0.9}$ \\
10 & $33.4^{+2.5}_{-1.3}$  & $30.7^{+5}_{-0.5}$ &  $26.8^{+0.7}_{-0.6}$\\ 

Average & 24 $\pm$ 2 & 27.2 $\pm$ 0.6 & 24.8 $\pm$ 0.6  \\
Average $m_{E}$ &  & -0.6 $\pm$ 0.2 & -0.14 $\pm$ 0.05 \\

\enddata
\label{tab:FWHM3}
\end{deluxetable}

\subsection{Fitting procedure}
For each choice of spectral index for the power law dependence of the
diffusion coefficient, $\mu$, we constructed a two dimensional grid in
the parameter space of $(B_0,\eta E_{h}^{1-\mu})$ for which we
calculated radial profiles at 0.7, 1, and 2 keV for each point in the
grid (recall that $\eta E_{h}^{1-\mu}$ is the constant scaling factor
of the diffusion coefficient, $D$). From this, we obtained both the
FWHM at 2 keV and the specific value of $m_E$ at 2 keV from
log(FWHM(2keV)/FWHM(1keV))/log(2). Thus we had numerical results for a
large number of discrete points over this parameter space and manually
found the point that simultaneously reproduced the values of both
$B_0$ and $\eta$ obtained from our observations.

We obtained the stated uncertainties by varying the parameters $\eta$
and $B_0$ around this best-fit value to find the domain in which the
observations were still satisfied within their respective
uncertainties. While not a formal error-analysis, this procedure is
adequate to put approximate lower and upper bounds on our estimates. 

\subsection{Loss-Limited Model}
The best-fit parameters are shown in Table~\ref{tab:num}, where
$\eta_2$ is the strength of the diffusion coefficient divided by the
Bohm diffusion coefficient at an energy of $\sqrt{\nu/c_m B_{0}}$,
which depends also on the fitted value for $B_{0}$. Note that this calculation included the convolution of the single particle emissivity with the solution of the continuous energy loss convection-diffusion equation for a spectrum of electrons exponentially cut off at the shock.

\begin{deluxetable}{c c c c c c c}
\tabletypesize{\scriptsize}
\tablecaption{Best fit parameters for the Filaments in varying values of $\mu$ (Numerical Results)}
\tablenotetext{*}{Results of fitting the data using our generalized diffusion model for the loss-limited case outlined in Section 3.  Dashes denote places where fits were unobtainable.  See note on Table~\ref{tab:anal} for $B_{0}$ and $\eta_2$}.
\tablewidth{0pt}
\tablehead{ &\multicolumn{2}{c}{Filament 1} & \multicolumn{2}{c}{Filament 2} & \multicolumn{2}{c}{Filament 3} }
  \startdata
$\mu$ &$\eta_{2}$ & $B_{0}$ &$\eta_{2}$ & $B_{0}$ & $\eta_{2}$ & $B_{0}$ \\ [.5ex]
0 & 7.5 $\pm$ 2 & 142 $\pm$ 5 & - & - & $\lapprox$ 0.1& 77 $\pm$ .8\\
1/3 & 4 $\pm$ 1.3 & 120 $\pm$ 5 & - & - & $\lapprox$ 0.1 & 76 $\pm$ 1.4 \\
1/2 & 3 $\pm$ 1.1 & 112 $\pm$ 4 & - & - & $\lapprox$ 0.1 & 75 $\pm$ 1.0 \\
1 & 2 $\pm$ 1.0 & 100 $\pm$ 3 & 22 $\pm$ 3 & 214 $\pm$ 4 & $\lapprox$ 0.1 & 74 $\pm$ 1.1 \\
1.5 & 1.9 $\pm$ 1.2 & 95 $\pm$ 3 & 9 $\pm$ 1.2 & 167 $\pm$ 4 & $\lapprox$ 0.1 & 74 $\pm$ 1.2   \\
2 & 2 $\pm$ 1.0 & 92 $\pm$ 4  & 7 $\pm$ 1.1 & 152 $\pm$ 4 & $\lapprox$ 0.1 & 73 $\pm$ 1.2 \\
\hline
\hline 
\\
& &\multicolumn{2}{c}{Filament 4} && \multicolumn{2}{c}{Filament 5} \\
\hline
\\
$\mu$ &&$\eta_{2}$ & $B_{0}$ &&$\eta_{2}$ & $B_{0}$\\
0 &&  $\lapprox$ 0.2 & 113 $\pm$ 2 && -& -\\
1/3 &&  $\lapprox$ 0.2 & 112 $\pm$2 && -& -\\
1/2 && $\lapprox$ 0.2  & 111 $\pm$2 && - & -\\
1 &&  $\lapprox$ 0.2 & 109 $\pm$ 2 && 80 $^{+\infty}_{-4}$   & 206 $\pm$ 3\\
1.5 && $\lapprox$ 0.2 & 108 $\pm$ 2 && 19 $\pm$ 2  & 140 $\pm$ 2\\
2 && $\lapprox$ 0.2 & 107 $\pm$ 2 && 12 $\pm$ 1.0  & 120 $\pm$ 2\\

\enddata
\label{tab:num}
\end{deluxetable}

\subsection{Magnetically Damped Model}

The magnetically damped model predicts that the data should show FHWMs
that are only weakly dependent on energy, caused by the electron
distribution's cut-off, with values of $m_{\rm E}$ on the order of
$-0.1$.  This is decidedly not what we observe (see Figure 9 and Tables 3-5), as the averaged
filaments all display values of $|m_{\rm E}| \gapprox 0.14$ at 2 keV
and $\gapprox 0.3$ at 1 keV.  This does not demonstrate that
post-shock magnetic field damping cannot occur, or that rims might not
be magnetically damped at much lower observation energies, but it
provides sufficient evidence that the damping length must at least be
large enough to be unimportant, i.e., larger than the synchrotron-loss
length, for electrons radiating at keV energies.  Therefore, we
confidently conclude that the X-ray rims of SN1006 are \emph{not} well
described by the magnetically damped model.

\section{Discussion}

In this section we will analyze the results of applying our model of
Section 3 to the SN1006 data presented in section 4.

In fitting the data for various values of $\mu$, we were able to
acquire the best fit diffusion coefficient energy relation in the
region in which electron energies are relevant for keV emission. 

Fitting the data allowed us to constrain both the magnitude and the
energy-dependence of the diffusion coefficient $D$, where the latter is reflected in the values of $\mu$ for which we could obtain fits.
However, fixing the magnitude of $D$ at some energy only fixes the
combination $\eta E_{h}^{1-\mu}$, so there is a degeneracy in the
choice of $\eta$ and $E_{h}$.  There are two restricting
conditions. The first is that $D(E)>D_{B}(E)$ at all energies, as the
Bohm coefficient is the minimum allowable. In other words,
\begin{equation}
\eta\left(\frac{E}{E_h}\right)^{\mu-1}>1.
\end{equation}
For $\mu>1$, this restricts $E^{\mu}$ diffusion to above some
threshold energy $E_{h}$, while for $\mu<1$, this restricts $E^{\mu}$
diffusion to below some maximum energy $E_h$. It also means that the
constant $\eta$ must always be greater than 1 as $D(E_h)/D_b(E_h)$ =
$\eta$. The second restriction is that $E_{h}$ should be outside the
relevant energy range for electrons emitting keV X-rays, as all of our
calculations had a fixed value for $\mu$.  What we can find, however,
is the minimum (for $\mu<1$) or maximum (for $\mu>1$) value of this
energy bound by fixing $\eta$ at 1. For our successful fits that have
non-negligible diffusion coefficients, the results of doing this give
values of $E_h$ that are so far outside the 0.7--7 keV photon
range that we can easily find an appropriate $E_h$ to satisfy the
above conditions.

From Tables~\ref{tab:anal} and~\ref{tab:num}, we see that several of
our averaged filaments require very small diffusion coefficients, well
below the Bohm value, even using the amplified magnetic field. This obviously implies that varying the
parameter $\mu$ will have no effect on the fits, as is evident in the
fitted values for $B_0$. Qualitatively, this means that the transport
of electrons is being carried out dominantly by the convection of
plasma away from the shock, and that each electron will stay attached
to its particular fluid element.  This result is required by the
strong energy dependence ($m_{\rm E} \sim -0.5$) of the filament
widths, as the presence of diffusion will always drive $m_{\rm E}$
towards 0 (or beyond to positive numbers in the case of $\mu>1$. On
the other hand, filaments with non-negligible diffusion coefficients
are all consistent with the condition that $D>D_{\rm Bohm}$, within
their respective uncertainties. This may suggest that some mechanism
is severely limiting electron diffusion in various regions of the
remnant, primarily the SW.  

One shortcoming of the magnetically damped model is the requirement
that $B(x)^{2}D$ is constant.  This implies that the diffusion
coefficient varies as $1/B(x)^2$, when we have explicitly written the
diffusion coefficient as proportional to $1/B(x)$ in our formalism.
However, this does not affect our conclusion that the magnetically
limited model is a poor fit to the data, as the qualitative behavior
of the FWHMs as a function of energy would be the same.  This
requirement presents no issues in the case of the loss-limited model,
as both $D$ and $B$ are spatially uniform in the narrow region behind
the shock.

Our finding that rim widths drop too rapidly with energy to allow
significant diffusion suggests the possibility of ``sub-Bohm
diffusion'' ($\lambda < r_g$, or $\eta < 1$) in astrophysical sources.
This possibility has important implications for acceleration times,
since much smaller diffusion coefficients $D$ would result in much
shorter acceleration times to a given energy.  There has been
considerable discussion of the possibility of sub-Bohm diffusion in
the literature.  For instance, Zank et al.~(2006) find that in
perpendicular shocks in the solar wind, effective mean free paths can
be an order of magnitude or more less than the gyroradius. They find
some supporting evidence in heliospheric observations.  Using a 3D
hybrid MHD-kinetic code, Reville \& Bell (2013) studied the
development of shock precursors generated by accelerated particles,
finding sub-Bohm behavior for both parallel and oblique shocks, but
more pronounced for parallel shocks.  However, some of their simulations find that the Bohm limit is still respected using the amplified magnetic field.  Reville \& Bell discuss other possible ambiguities in the definition of the Bohm limit, including the possibility of highly inhomogeneous magnetic fields on small scales. At any rate, it seems clear that the
complexities of the propagation of particles in the presence of
dynamic self-generated magnetic turbulence are such that the
suppression of diffusion to levels considerably below those implied by
the Bohm limit is not ruled out by theoretical considerations.  We
emphasize that in spite of the elaborate theoretical structure we have
presented, our limits on the diffusion coefficient are closely related
to the rapid drop in filament widths with photon energy that we see in Figure \ref{widthdata}.
The rate of shrinkage is too large to tolerate much particle
diffusion, independently of detailed modeling.  However, detailed
quantitative statements are dependent on the details of our mechanism
for fitting filament widths, and on inevitable projection and
curvature effects.  Our models do predict considerably thinner
filaments at 4 keV than at 2; while our current observations do not
have adequate photon statistics to test this prediction, future
studies should be performed to clarify this important issue.

Finally, we call attention to the $\mu$ dependence in the electron cut-off energy used in our model as described in Equation (19).  For a synchrotron emitting source, this cut-off energy corresponds to a rolloff frequency of $\nu_{roll} \propto E_{cut}^2 B_0 \propto B_{0}^{(\mu-1)/(\mu+1)} v_s^{4/({\mu+1})}$.  If $\mu=1$ as in the standard Bohm assumption, we find a rolloff frequency that is independent of the magnetic field and solely a function of the shock speed. In that case, we would expect constant rolloff frequencies along the same filaments and only a relatively weak azimuthal dependence, predictable from observed proper motions.  On the other hand, if $\mu \neq 1$, then we recover a $B_0$ dependence, which could account for the some of the systematic order-of-magnitude azimuthal variation of the measured rolloff frequencies seen in both SN1006 (Katsuda et al. 2010, Miceli et al. 2009, Reynolds et al. 2012) and G1.9+0.3 (Reynolds et al. 2009).  For $\mu=2$, one would require a very large variation in $B_0$, largest at the brightness maxima, to explain the observed factor of 10 range in rolloff frequency, however.  

\subsection{Comparisons to Cas A}

A detailed application of our results to other SNRs such as Cas A will
require much more extensive analysis, but we can use the published
filament widths of Araya et al. (2010) for Cas A to get preliminary
estimates of the magnetic field strength and diffusion coefficient by
applying our model. In their data, it appears that the filaments in
Cas A shrink by a factor of $\sim 0.8$ between 0.3 and 3 keV, while the
filament widths appear to be energy-independent between 3 and 6 keV.
Qualitatively, this is consistent with the loss-limited model, as our
parameter $m_{\rm E}$ is predicted to decrease with energy.  For the
lower energy range of 0.3--3 keV, reproducing $m_{\rm E}\sim -0.1$
(equivalent to the factor of 0.8 drop in size) requires magnetic
fields on the order of 200-500 $\mu$G and diffusion coefficients about
$5 \times D_{\rm Bohm}$(3 keV), about an order of magnitude higher
than the values one obtains by neglecting the energy dependence. One
can also see directly from Figure~\ref{Etagraph} that $\mu < 1$ models
of the diffusion coefficient are excluded for $m_{\rm E}\sim -0.1$.

\section{Summary and Conclusions}

We have outlined a generalized diffusion model that solves the
continuous energy-loss convection-diffusion equation for electrons
subject to both convection and diffusion as they travel away from the
shock.  This model is able to incorporate arbitrary power-law energy
dependence of the diffusion coefficient $D$, in the form of $D \propto
E^\mu$, as well as arbitrary spatial dependence of the magnetic field
strength.  Assuming spherical symmetry, we then convolved this
electron distribution with the single electron power spectrum to
obtain a non-thermal emissivity, which we integrated along lines of
sight to obtain the specific intensity of the source as a function of
radial distance from its center. We specialized this model into two
general categories: a ``Magnetically Damped Model,'' which assumes
$B\propto \exp^{-x/a_b}$ for $x$ defined as the distance behind the
shock, and a ``Loss-limited Model'' which assumes a constant
post-shock magnetic field strength. Furthermore, we selected specific
values of the parameter $\mu$ (namely 0,1/3, 1/2, 1, 1.5, and 2) to
analyze.  We found that independent of model details, magnetically
damped models predict rim widths almost independent of photon energy,
while loss-limited models predict rim widths to shrink with increasing
photon energy at a rate dependent on the diffusion coefficient.
Our quantitative results are summarized in Figure~\ref{Etagraph}. 

With this model as our guide, we used \emph{Chandra} observations of
SN 1006 to measure the energy dependence of the thin, non-thermal rims
in the NE and SW quadrants.  For the SW, these are the first such
measurements, utilizing data from a recent Large \emph{Chandra}
Project.  The SW filament profiles are of similar width to those in
the NE.  This is consistent with the results of Winkler et al.~(2014),
who find the conditions in both regions to be similar. Furthermore,
the filament widths of SN1006 show a decrease with photon energy,
which we have shown has important physical consequences for both the
diffusion coefficient and the post-shock magnetic field, and is
incompatible with a magnetic damping model.  In the other SNRs for
which magnetic fields have been inferred from rim thicknesses, the
energy dependence of the widths should be examined similarly, as
evidenced by our quick comparison with the results of Araya et
al.~(2010) for Cas A.

Using our generalized diffusion model and its subsets outlined above,
we find the measured widths of the SW filaments of SN 1006, like those
previously reported for the NE, favor magnetic fields on the order of
100 $\mu$G, significantly amplified above the typical interstellar
medium value of about 3 $\mu G$.  The strength of our model is that it
encompasses all effects included by previous authors in this type of
investigation.
 
We also conclude that for the filaments of SN 1006 , and even the
filaments of Cas A, values of $\mu<1$ are, for the most part, less
able to reproduce the data.  This is due to the result that the lowest
possible $m_{\rm E}$ for a given diffusion coefficient is $(\mu
-1)/4$, which requires $\mu \gapprox 1$ in Filaments 2 and 5 for SN
1006 and the majority of filaments in Cas A (Araya et al. 2010).
While not conclusive, this result is in agreement with the
calculations of Reynolds (2004), which predict SNR images that do not
resemble observed SNRs for $\mu<1$.  Thus both Kolmogorov turbulence
($\mu = 1/3$) and Kraichnan ($\mu = 1/2$) are disfavored.  In an
application to other SNRs, values of $\mu>1$ allow completely
energy-independent filament widths even with a cut-off in the injected
spectrum of electrons.  Thus, the results of Araya et al.~(2010), who
found energy independent FWHMs in Cas A between 3-6 keV, could result
from strong diffusion and a very high electron cut-off energy, or from
non-Bohm-like diffusion with $\mu>1$.  In the latter case, our model also predicts that the rolloff frequency should depend on magnetic field strength and thus could have significant azimuthal dependence.

We also find that that the magnitudes of the diffusion coefficients in
the filaments of SN1006 are split into two distinct categories.  One
group, due to the strong energy dependence of their filament widths,
requires negligible amounts of diffusion (i.e. less than the Bohm limit) in order to reproduce the
observations.  The occurrence of sub-Bohm diffusion is thus far undocumented and would be a groundbreaking result. However, there are inherent uncertainties in our measurements of the FWHMs due to projection, overlap, and averaging effects which may have influenced our numerical calculations.  Thus, we do not claim strong evidence of this theoretically hard to explain phenomenon and note that further study is needed.   On the other hand, we are much more confident in the rest of our conclusions, which are fairly robust and do not depend on sub-Bohm diffusion. The second group, with less energy-dependent filament
widths, is consistent with diffusion coefficients close to but above the
Bohm limit. 
 This result is of
interest because the strength of diffusion is directly tied to the
maximum energy attainable by electrons being accelerated at the shock
front. Diffusion coefficients much larger than the Bohm value would
have suggested weaker scattering, which in turn would reduce the
maximum energy (Reynolds 1998).  However, in our model of the
filaments, the radial profiles are produced by electrons in the
post-shock region, so pre-shock electrons could have much different
diffusive properties.  Our fits predict continuing shrinkage of
filament widths at higher energies than 2 keV, though photon
statistics in our current observations are not adequate to test this.
(Our 2--7 keV band is dominated by photons near 2 keV.)  A longer
observation of the SW region of SN 1006 with {\sl Chandra} could allow
division of that band into 2 -- 4 and 4 -- 7 keV bands, permitting
this important test.  Our surprising result of rapid shrinkage of some filaments requiring sub-Bohm diffusion coefficients can be searched for in other thin-rim remnants such as Tycho.  

Finally, we find that the results of applying our generalized
diffusion model are remarkably consistent with the results obtained by
simply fitting Equation 6 to the data (recall that Equation 6 was the result of applying the $\delta$-function approximation for the electron spectra to the catastrophic dump version of the convection-diffusion equation).  This is in spite of the fact
that our model solves the continuous energy loss convection-diffusion
equation for the electron distribution, uses the full synchrotron
emissivity, and includes a cutoff in the injected spectrum of
electrons, all of which we have shown to have important effects on the
FWHMs and their energy dependence.  This may simply be a unique result
for the observational data from SN1006, or it could suggest that the
the effects of adding each of these more detailed considerations
cancel each other out when combined. On the other hand, Equation 6 is
incapable of describing magnetically damped filaments, which may occur
in other remnants (e.g., Marcowith \& Casse 2010), though we have
ruled this out for SN1006.

The general formalism presented here is applicable to the thin,
non-thermal filaments observed in nearly all historical SNRs, and has
the potential to provide a consistent estimate of magnetic-field
amplification across the variety of ambient environments into which
these remnants are expanding.

Support for this work was provided by the National Aeronautics and
Space Administration through {\it Chandra} Grant Number GO2-13066, issued
by the {\it Chandra} X-ray Observatory Center, which is operated by the
Smithsonian Astrophysical Observatory for and on behalf of NASA under
contract NAS8-03060\@.  

We thank the anonymous referee for an extremely careful reading of this paper, and for suggestions that have led to substantial improvements.

\appendix

\section{Appendix}

We summarize here some details of the relation of our models to
observations.  Our basic conclusion is that the variation of filament
widths with energy contains essential information required to compare
models, and to obtain quantitative estimates of the magnetic field and
diffusion coefficient.  Without this information, there are several
competing models that can allow for a wide range of magnetic-field
strengths and diffusion coefficients for the same filament width.

We assume throughout a spherical shock surface, in which the peak of
synchrotron emission occurs at a radius slightly behind the shock due
to the geometry of the line of sight integration. If instead a plane
shock with velocity exactly in the plane of the sky is assumed,
derived quantities will vary somewhat.  In addition, we find that the
$\delta$-function approximation gives a poor representation of the
spatial distribution of high-energy electrons, resulting in an
underestimate of filament widths.  It is also essential to consider
the cutoff in the electron spectrum above some maximum energy as it
causes filament widths to shrink with photon energy in a
model-independent way.  At photon energies close to the rolloff
frequency, the only way to have truly energy-independent rim sizes is
with $\mu>1.$ In energy-loss models with $\mu\leq 1$, or damping
models, $|m_{\rm E}|$ will always be at least $\sim$ 0.1 if the cutoff
is not exceptionally high (well above the keV band).

Then the strength of the energy-dependence of filament widths serves
as the essential discriminant among models.  If a weak energy
dependence ($0.1 \lapprox |m_{\rm E}| \lapprox 0.2$) is observed for
photon energies near the synchrotron rolloff frequency, the behavior
at lower photon energies should be inspected as it will have greater
discriminatory power.  Here ``lower energies'' means energies lower
than those where diffusion and the electron cutoff start to become
important, which depend on the source parameters.  Near the rolloff
frequency, many effects can combine to cause weak energy-dependence
of filament widths.

If moderately strong energy-dependence of filament widths is observed
($0.2 \lapprox |m_{\rm E}| \lapprox 0.5$) at a specific energy, then a
magnetic-field damping model can be ruled out at that energy and
above.  This is the region in which it can be assumed that diffusion
is important in competing with advection. The details of this then
depend on the assumed model of diffusion.  And if very strong
energy-dependence of filament widths is observed ($|m_{\rm
E}| \gapprox .5$) then the only explanation is weak diffusion and the
predominance of advection as the electron transport mechanism.

Finally, if filaments widths are ever observed to be growing with
energy then the only known explanation would be $\mu>1$ diffusion.
Higher values of $\mu$ allow for more rapid changes in $m_{E}$ as a
function of energy.




\begin{deluxetable}{c c c}
\tablecaption{Symbol Glossary (Numerical values in CGS)}
\tablewidth{0pt}
\tablehead{\colhead{Symbol} & \colhead{Expression/Value}& \colhead{Explanation}}
\tablenotetext{a}{Katsuda et al. 2009}
\tablenotetext{b}{Calculated using the angular size in Green's catalog (2009) and the distance to the remnant given by Winkler et al. (2003)}
\startdata
\multicolumn{3}{c}{Diffusion Coefficient, D = $\eta D_{
\rm B}(E_h)\left(\frac{E}{E_h}\right)^{\mu}$}\\

$D_{B}$ & $\frac{C_{\rm d}E}{B}$ & ``Bohm-limit'' of D\\
 $C_{\rm d}$ & 2.083 $\times$ $10^{19}$ \\
 $\eta$ & & Scaling factor of D\\
 $\mu$ & & Power index for D with E\\
  $E_h$ & & Arbitrary energy (keeps $\eta$ dimensionless)\\
 $\eta_2$& $\frac{D}{D_{\rm B}}$(2 keV) & Relative D at $h\nu$=2 keV \\
  \hline
  \multicolumn{3}{c}{Synchrotron Parameters}\\
  
  $\tau_{synch}$ & $\frac{1}{bB^2E}$ & Synchrotron lifetime\\ 
  $l_{ad}$ & $\frac{v_s}{4}\tau_{synch}$& Advective length\\
  $l_{diff}$ & $\sqrt{D\tau_{synch}}$& Diffusive length\\
  
$\nu_m$ & $c_m E^2 B$ & $\delta$-function synchrotron frequency \\
 $c_{m}$& 1.82 $\times$ $10^{18}$ \\
 $c_{1}$ & 6.27 $\times$ $10^{18}$ \\
 b & 1.57$\times$ $10^{-3}$\\
 \hline
   \multicolumn{3}{c}{Diffusion Model Parameters}\\

$B_0$ & & Immediate Post-shock B-field\\
$x_{1/2}$ & & FWHM of radial intensity profile \\
$m_{\rm E}$ & $\frac{\partial\log(x_{1/2})}{\partial\log(E_\gamma)}$& Power index of $x_{1/2}$ with $E_\gamma$ \\
f(x,E) & & $e^-$ spatial and energy distribution\\
$a_b$ & & Length scale in B-damping model\\

 \hline
   \multicolumn{3}{c}{SN1006 Parameters}\\

 $v_{s}^{\rm a}$  & 5 $\times$ $10^{8}$ & Shock Velocity\\
 $r_{s}^{\rm b}$ & 2.96 $\times$ $10^{19}$  &Shock Radius \\
 s & 2.2 & $e^-$ spectral index\\
\enddata
\label{tab:const}
\end{deluxetable}

\end{document}